\documentclass[letterpaper,11pt]{article}
\usepackage[DIV12]{typearea}
\usepackage{longtable}
\usepackage{booktabs,colortbl}
\usepackage{multirow}
\usepackage{amsmath}
\usepackage{amssymb}
\usepackage{bbm}
\usepackage[utf8]{inputenc}
\usepackage{pdflscape}
\usepackage{xspace}
\usepackage[caption=false]{subfig}
\usepackage{nicefrac}
\usepackage{graphicx}
\usepackage{cite}  
\usepackage[small]{caption2}
\usepackage{mathtools}
\usepackage{nccfoots}
\usepackage{tikz}
\usetikzlibrary{calc,positioning}
\usepackage{listings}
\usepackage{cancel}
\usepackage{soul}
\usepackage{pdflscape}
\usepackage{rotating}
\usepackage{geometry}
\usepackage{bold-extra}

\lstdefinestyle{consola}
{backgroundcolor=\color{gray97},
basicstyle=\small\color{black}\ttfamily,
breaklines=true,
inputencoding=utf8,
extendedchars=true,
literate={á}{{\'a}}1 {ñ}{{\~n}}1 {ó}{{\'o}}1 {í}{{\'{\i}}}1 {é}{{\'e}}1 {ú}{{\'u}}1,
}
\lstdefinestyle{pyt}
{language=Python,
frame=Ltb,
framerule=0pt,
aboveskip=0.5cm,
framextopmargin=3pt,
framexbottommargin=3pt,
framexleftmargin=0.4cm,
framesep=0pt,
rulesep=.4pt,
backgroundcolor=\color{gray97},
rulesepcolor=\color{black},
stringstyle=\ttfamily,
showstringspaces = false,
basicstyle=\footnotesize\ttfamily,
commentstyle=\color{gray45},
keywordstyle=\color{black}\bfseries,
numbers=left,
numbersep=15pt,
numberstyle=\tiny,
numberfirstline = false,
breaklines=true,
inputencoding=utf8,
extendedchars=true,
literate={á}{{\'a}}1 {ñ}{{\~n}}1 {ó}{{\'o}}1 {í}{{\'{\i}}}1 {é}{{\'e}}1 {ú}{{\'u}}1,
}
\usepackage{color}
\definecolor{DarkGrey}{rgb}{0.1,0.1,0.1}
\definecolor{gray97}{gray}{.97}
\definecolor{gray75}{gray}{.75}
\definecolor{gray45}{gray}{.45}
\makeatother

\usepackage[all,cmtip]{xy}
\newlength{\xywd}
\newcommand{\xyrightarrow}[2][]{%
  \sbox{0}{$\scriptstyle#1$}%
  \xywd=\wd0
  \sbox{0}{$\scriptstyle#2$}%
  \ifdim\wd0>\xywd \xywd=\wd0 \fi
  \xymatrix@C\dimexpr\xywd+1em\relax{{}\ar[r]^{#2}_{#1}&{}}%
}

\renewcommand{\thefootnote}{\alph{footnote}}

\newcommand{\rep}[1]{\ensuremath\boldsymbol{#1}}
\newcommand{\crep}[1]{\ensuremath\bar{\boldsymbol{#1}}}
\newcommand{\Z}[1]{\ensuremath{\mathbb{Z}_{#1}}} 

\newcommand{\SU}[1]{\ensuremath{\mathrm{SU}(#1)}}

\newcommand{\U}[1]{\ensuremath{\mathrm{U}(#1)}}
\newcommand{\E}[1]{\ensuremath{\mathrm{E}_{#1}}}

\newcommand{\x}{\ensuremath{\times}}

\usepackage{xcolor}
\definecolor{darkgreen}{HTML}{109930}
\definecolor{pink}{rgb}{0.858, 0.188, 0.478}

\advance \headheight by 3.0truept       
\usepackage[pdftex,hidelinks,colorlinks=true,citecolor=darkgreen]{hyperref}
\hypersetup{
    pdftitle = {An autoencoder for heterotic orbifolds with arbitrary geometry},
    pdfauthor = {Escalante-Notario, Portillo-Castillo, Ramos-Sanchez}
}

\usepackage[toc,page]{appendix}    
\usepackage{cleveref}

\begin{document}

\begin{titlepage}

\begin{flushright}
\end{flushright}

\vspace*{1.0cm}

\begin{center}
{\Large\textbf{An autoencoder for heterotic orbifolds with arbitrary geometry}}

\vspace{1cm}
\textbf{Enrique Escalante--Notario$^{1,}$\footnote{\texttt{eescalante@estudiantes.fisica.unam.mx}}},
\textbf{Ignacio Portillo--Castillo$^{1,2,}$\footnote{\texttt{iportillo@uach.mx}}} \\ 
and
\textbf{Sa\'ul Ramos--S\'anchez$^{1,}$\footnote{\texttt{ramos@fisica.unam.mx}}}\\[5mm]

\textit{\small ${}^1$ Instituto de F\'isica, Universidad Nacional Aut\'onoma de M\'exico,\\ POB 20-364, Cd.Mx. 01000, M\'exico}\\
\textit{\small ${}^2$ Facultad de Ingenier\'ia Universidad Aut\'onoma de Chihuahua,\\Nuevo Campus Universitario, Chihuahua 31125, M\'exico.}
\end{center}

\vspace{1cm}

\vspace*{1.0cm}

\begin{abstract}
Artificial neural networks can be an important tool to improve the search 
for admissible string compactifications and characterize them. In this paper we 
construct the \texttt{heterotic orbiencoder}, a general deep autoencoder to study 
heterotic orbifold models arising from various Abelian orbifold geometries. Our 
neural network can be easily trained to successfully encode the large parameter
space of many orbifold geometries simultaneously, independently of the statistical
dissimilarities of their training features. In particular, we show that our 
autoencoder is capable of compressing with good accuracy the large parameter space 
of two promising orbifold geometries in just three parameters. Further, most 
orbifold models with phenomenologically appealing features appear in bounded regions 
of this small space. Our results hint towards a possible simplification
of the classification of (promising) heterotic orbifold models.
\end{abstract}

\end{titlepage}

\newpage
\setcounter{footnote}{0} 
\renewcommand{\thefootnote}{\arabic{footnote}}

\section{Introduction}

Compactifying the six extra dimensions of any of the string theories yields a huge
number of four-dimensional (4D) models. Recent scans include compactifications based 
on intersecting D-branes~\cite{He:2021kbj,He:2021gug,Loges:2022mao}, 
fermionic constructions~\cite{Senechal:1989me,Dienes:1990ij,Dienes:2006ut}, 
orientifolds of Gepner models~\cite{Dijkstra:2004cc,Anastasopoulos:2006da}, 
Calabi-Yau manifolds~\cite{Kreuzer:1998vb,Kreuzer:2000xy,Anderson:2011ns,Gray:2012jy,Scholler:2018apc}
and supersymmetric~\cite{Lebedev:2006kn,Lebedev:2007hv,Lebedev:2008un,Nilles:2014owa,Olguin-Trejo:2018wpw,Parr:2019bta} 
and non-supersymmetric~\cite{Blaszczyk:2014qoa,Perez-Martinez:2021zjj} 
heterotic orbifolds~\cite{Dixon:1985jw,Dixon:1986jc}. Many of those constructions 
include three generations\footnote{Interestingly, models with a different number of 
generations can be phenomenologically relevant due to strong dynamics available at 
low energies in these models~\cite{Ramos-Sanchez:2021woq}.} of the standard model (SM), 
its minimal supersymmetric extension (MSSM) or their completions in grand unified theories (GUT). 
However, most models are not of this kind, i.e.\ they display features that are not compatible 
with observations.

The properties of the resulting 4D theories depend on the choice of the geometry 
of the compact space, which is characterized by some selectable parameters. Their quantity and
admissible values are determined by the type of string compactification and strict consistency
constraints, such as modular invariance and tadpole cancelation.
The parameters and their possible values build enormous convoluted multidimensional spaces. 
Hence, the task of exploring all points in such parameter spaces yielding admissible 
models and, then, identifying among them the best candidates to describe nature becomes 
insurmountable or, in the best of cases, extremely challenging.
Despite being time-consuming, random searches have shown certain 
efficacy~\cite{Denef:2004ze,Giryavets:2004zr,Gmeiner:2005vz,Douglas:2006xy,Abel:2014xta,Cole:2019enn},
but we may profit from a more ingenious approach.

The latest developments of numerical methods and, specifically, of machine learning (ML) techniques
for information processing along with the accessibility to faster computational hardware, has favored 
the applications of ML in various fields, including string phenomenology~\cite{Ruehle:2020jrk}. 
Basic numeric applications in this endeavor, such as genetic algorithms~\cite{Damian:2013dq,Damian:2013dwa,Cole:2019enn,Loges:2021hvn},
have led to more complex ML string-phenomenology setups based on
decision trees~\cite{Carifio:2017bov,Ashmore:2019wzb,Parr:2019bta}, 
deep neural networks~\cite{He:2017aed,He:2019vsj,Demirtas:2020dbm,Berglund:2021ztg,Jejjala:2022lxh}, 
support vector machines~\cite{Bull:2018uow,Erbin:2020tks}, 
reinforcement learning~\cite{Halverson:2019tkf,Constantin:2021for,Abel:2021rrj}, 
autoencoders~\cite{Mutter:2018sra,Otsuka:2020nsk,Deen:2020dlf}, and several 
others~\cite{Halverson:2020opj,CaboBizet:2020cse,Damian:2022xfk}.

In particular, autoencoders~\cite{baldi2012autoencoders} are artificial neural networks (NN)
frequently used to furnish a compressed representation of some multidimensional input data.
By using them to reduce the dimensionality of their parameter spaces,
they might become instrumental in the classification of string models and the 
identification of the most phenomenologically suitable ones.

We focus here on (Abelian toroidal) orbifold compactifications of the \E8\x\E8 heterotic
string in the bosonic formulation. In their simplest form, they correspond to assuming that 
the compact space adopts the structure given by the quotient $\mathbbm R^6/S$, 
where $S$ denotes any of the Abelian space groups classified in ref.~\cite{Fischer:2012qj}. 
Once $S$ and thereby the orbifold geometry is chosen, modular invariance demands the orbifold 
to be embedded in the gauge degrees of freedom of \E8\x\E8, which can be done via eight 
16-dimensional (16D) rational vectors known as shifts and Wilson lines. These 128 rational 
parameters are then subject to additional modular invariance conditions~\cite{Ploger:2007iq}, which
ensure anomaly freedom of the resulting 4D effective theory. The exploration of this 128-dimensional
parameter space has been automatized in the \texttt{orbifolder}~\cite{Nilles:2011aj}. However, 
it is known that not all models can be found with this software~\cite{Parr:2019bta}.
In ref.~\cite{Mutter:2018sra} it has been shown for one particular $S$ that an autoencoder
may pack with acceptable accuracy the properties of an orbifold in a two-dimensional 
latent space.

One of our goals here is to extend the results of ref.~\cite{Mutter:2018sra}, to achieve a
tool capable of analyzing various space groups at the same time, aiming at a general classification 
of all orbifolds in a simplified latent space. This poses a challenge as models arising 
from different orbifold geometries exhibit very different patterns in their parameter 
spaces. Additionally, our tool should achieve a better accuracy than previous efforts 
and yield a better understanding of how promising compactifications are accommodated 
in the latent space. With this aim, we design a (fully connected, symmetric) deep autoencoder,
which we call \texttt{heterotic orbiencoder}, that can receive multiple orbifold 
geometries and encode their information in a tridimensional (3D) latent space. 

Our work is organized as follows. After a noncomprehensive review on heterotic orbifolds
in section~\ref{sec:orbifolds}, we devote our section~\ref{sec:OptimizingConfigurations}
to exploring a variety of architectures and function configurations (combinations of optimizer, 
activation and loss functions) in order to identify the optimal structure of our deep 
autoencoder. These findings allow us develop the \texttt{heterotic orbiencoder}, whose
properties and commands are described in section~\ref{sec:HOE}. 
We exemplify its power by applying it on two phenomenologically appealing 
orbifold geometries (\Z8--I (1,1)~\cite{Nibbelink:2013lua,Olguin-Trejo:2019hxk}
and \Z{12}--I (1,1)~\cite{Kim:2006hw,Kim:2007mt,Huh:2009nh}) in section~\ref{sec:TwoGeometries}. Finally, we discuss our 
main observations in section~\ref{sec:conclusions}. In our appendix we discuss some
challenges we face to optimize the speed and accuracy of the autoencoder while avoiding 
{\it overfitting}.

\section{Features of heterotic orbifolds}
\label{sec:orbifolds}

In order to fix our notation, let us introduce some basics on heterotic orbifold compactifications.
(For more detailed discussions, see e.g.~\cite{Bailin:1999nk,Ramos-Sanchez:2008nwx,Vaudrevange:2008sm,Choi:2020dws}
or~\cite{Olguin-Trejo:2018wpw}.)
The space group $S$ of a six-dimensional (6D) orbifold $\mathbbm R^6/S$ leading to $\mathcal N=1$ supersymmetric
effective field theories at low energies can be specified by up to two generators,
$\vartheta,\omega$, of the orbifold rotational subgroup, 
and six basis vectors $e_\alpha$, $\alpha=1,\ldots,6$, of the lattice of a $\mathbbm T^6$ torus.
An arbitrary element of $S$ can be written as $(\vartheta^k\omega^\ell, m_\alpha e_\alpha)\in S$,
where $0\leq k<N$ and $0\leq \ell<M$, $k,\ell\in\mathbbm N$, such that $N$ and $M$ denote the
orders of $\vartheta$ and $\omega$, respectively. Further, $m_\alpha\in\Z{}$ only in the absence 
of roto-translations. It is convenient to 
identify the conjugacy classes of $S$, such that we can build a set of inequivalent space-group
elements by choosing one representative from each conjugacy class.

Orbifolds have curvature singularities also known as fixed points, since they are left invariant 
under space group elements. Each singularity can be associated with a conjugacy class of $S$ 
exhibiting nontrivial rotational action. This implies that the number of fixed points $n_\mathrm{fp}$
depends on $S$, i.e.\ $n_\mathrm{fp}=n_\mathrm{fp}(S)$. In order to better identify the fixed points, it is 
customary to split the conjugacy-class representatives into {\it sectors} labeled by
the pairs $(k,\ell)$. A fixed point $x_f\in\mathbbm R^6$ is said to belong to the $(k,\ell)$ 
sector if there exists a conjugacy-class representative $(\vartheta^k\omega^\ell, m_\alpha e_\alpha)\in S$, 
such that
\begin{equation}
 (\vartheta^k\omega^\ell, m_\alpha e_\alpha) x_f ~=~
    \vartheta^k\omega^\ell x_f + m_\alpha e_\alpha ~\stackrel{!}{=}~ x_f\,.
\end{equation}

The next building block to obtain a heterotic orbifold is the embedding of $S$ into the 16D gauge degrees
of freedom associated with \E8\x\E8. Both the six toroidal shifts $e_\alpha$ and the two 
rotational generators of $S$ can be embedded as 16D rational gauge shift vectors. In our 
conventions, the gauge embedding is given by 
\begin{equation}
\label{eq:gaugeEmbedding}
 e_\alpha ~\hookrightarrow~ V_\alpha \quad\text{with }\alpha=1,\ldots,6\,,\qquad 
 \vartheta~\hookrightarrow~ V_7\qquad\text{and}\qquad 
 \omega   ~\hookrightarrow~ V_8\,.
\end{equation}
In this embedding the vectors $V_A$, $A=1,\ldots,8$, encode the orbifold action on the 
\E8\x\E8 space as shifts in the gauge momenta. $V_\alpha$ are frequently called Wilson lines.

The gauge shift vectors $V_A$ must comply with a number of consistency 
conditions. For example, since $V_7$ ($V_8$) is a gauge embedding of the rotation 
$\vartheta$ ($\omega$) of order $N$ ($M$), then it must fulfill that $N V_7$ ($M V_8$)
be in the root lattice of \E8\x\E8. Hence, $N$ and $M$ are also known as the orders
of $V_7$ and $V_8$, respectively. The relations among the various $e_\alpha$ due to 
the actions of $\vartheta$ and $\omega$ on them, are translated as relations among 
the different $V_A$, $A=\alpha=1,\ldots,6$, that must be fulfilled, and also establish the 
order $N_A$ of these shift vectors. Additionally, these shift vectors must satisfy
certain modular invariance conditions~\cite{Ploger:2007iq}.

For each choice of $S$, there is a large (and unknown) number of inequivalent gauge embeddings $V_A$ 
that fulfill all consistency conditions, and lead to admissible 4D effective field 
theories. This vast set of effective theories can be regarded as the landscape of heterotic orbifold models.
These models exhibit various gauge groups $G_\mathrm{4D}\subset\E8\x\E8$ of rank 16 
and (massless and massive) matter fields building representations under $G_\mathrm{4D}$. 
One can focus only on the massless states because the fundamental scale of these 
constructions is the string scale, which is close to the Planck scale and hence
decoupled from observable physics.

The massless matter spectrum is given by the closed strings that are invariant 
under the orbifold. They can be split into i) strings that are closed (and massless)
prior to the orbifold action, known as {\it untwisted} strings, and ii) strings that
close only due to the action of the orbifold, called {\it twisted} strings. In fact,
the representatives of the conjugacy classes of $S$ indicate how strings transform
at the fixed points and, thus, are used to establish the boundary conditions of the 
strings to close. In this scheme, untwisted strings are associated with the 
$(k,\ell)=(0,0)$ sector, as neither $\vartheta$ nor $\omega$ twists the strings 
to close in this case. Twisted strings arise from the $(k,\ell)\neq(0,0)$ twisted
sectors. We focus here on untwisted strings because they yield the gauge group,
which is considered in this work to be the key property of the model.\footnote{Notice 
that this is an oversimplification, whose effects must be considered when evaluating
our results.}

Untwisted matter states carry the gauge momenta $p$ associated with the original 
\E8\x\E8 gauge bosons of the heterotic string in 10D that are left invariant 
under the gauge embedding of the orbifold. The unbroken gauge group $G_\mathrm{4D}$ can be obtained from 
requesting that the momenta $p$ be invariant under the gauge shifts $V_A$, which
amounts to demanding that
\begin{equation}
\label{eq:G4Dcondition}
  p\cdot V_A ~=~ 0\mod 1\quad\forall A \quad\iff\quad 
  p^{(1)}\cdot V_A^{(1)} ~=~ 0\mod 1 ~=~ p^{(2)}\cdot V_A^{(2)}\qquad \forall A\,,
\end{equation}
where, for future convenience, we have separated the two \E8 components, such 
that each $p^{(a)}$, $a=1,2$, and $N_A V_A^{(a)}$, $A=1,\ldots,8$, are elements 
of the eight-dimensional (8D) root lattice of \E8. The \E8 momenta can be written as
\begin{equation}
\label{eq:E8momenta}
 p^{(a)} ~\in~ \left\{
 \begin{array}{lcl}
 (\underline{\pm1,\pm1,0^6})\,,      && \\
 \left( (\pm\nicefrac12)^8\right)\,, && \text{even \# of $+$}\,,
 \end{array}
 \right.
\end{equation}
where the power in the entries denotes repetition, and the underscore 
stands for all possible permutations. 

The subset $I_0$ of the momenta~\eqref{eq:E8momenta} satisfying eqs.~\eqref{eq:G4Dcondition}
builds $G_\mathrm{4D}$, and clearly depends on the gauge embedding vectors $V_A$. It will
be convenient to define the subgroups $G_\mathrm{4D}^{(a)}$, such that 
$G_\mathrm{4D}=G_\mathrm{4D}^{(1)}\x G_\mathrm{4D}^{(2)}$, which are built by the invariant 
\E8 simple roots in the subsets
\begin{equation}
\label{eq:I0}
   I_0^{(a)}~:=~\left\{ p^{(a)}  ~\big|~ V_A^{(a)}\cdot p^{(a)}~=~0\mod 1, A=1,\ldots,8\right\}\,,\qquad a=1,2\,.
\end{equation}
Note that the same $G_\mathrm{4D}^{(a)}$
can be generated for different (although equivalent) sets $I_0^{(a)}$.
This is related to the symmetries of the \E8 root lattice, which include
\E8 Weyl rotations and lattice translations, among other transformations, see e.g.~\cite{GrootNibbelink:2017usl}. 
Modding out these symmetries is not feasible at the moment. However, observe that 
the number of invariant roots $N_0^{(a)}:=\big|I_0^{(a)}\big|$ must be an invariant 
under all symmetry transformations, and can hence be used as a characteristic feature
of a heterotic orbifold model.

Interestingly, at the fixed points of the twisted sectors, the closed 
strings related with the gauge bosons do not perceive the whole action 
of the orbifold, but only its local action. This is encoded in the 
conjugacy-class representative 
$(\vartheta^k\omega^\ell, m_\alpha e_\alpha)\in S$ that defines each fixed 
point. By using eq.~\eqref{eq:gaugeEmbedding}, we see that the shift
vector $V_\mathrm{local}$ associated with the localized action of
the orbifold at such fixed point is given by
\begin{equation}
\label{eq:gaugeEmbeddingTwisted}
 (\vartheta^k\omega^\ell, m_\alpha e_\alpha)~\hookrightarrow~
 V_\mathrm{local}:=k V_7 + \ell V_8 + m_\alpha V_\alpha\,.
\end{equation}
Hence, the local gauge symmetry group $G_{\mathrm{local},n}$ realized at the 
$n$th singularity is built by the momenta in the sets
\begin{equation}
\label{eq:In}
   I_n^{(a)}~:=~\left\{ p^{(a)} ~\big|~ V_{\mathrm{local},n}^{(a)}\cdot p^{(a)}~=~0\mod 1 \right\}\,,\qquad a=1,2\,,\quad n=1,\ldots,n_\mathrm{fp}\,.
\end{equation}
As for the gauge group $G_\mathrm{4D}$ associated with the untwisted sector, the cardinalities 
$N_n^{(a)}:=\big|I_n^{(a)}\big|$ are invariant properties of the local gauge 
group $G_{\mathrm{local},n}$.

\section{Optimizing an autoencoder for heterotic orbifolds}
\label{sec:OptimizingConfigurations}

\subsection{Orbifold datasets}
\label{sec:dataset}

Considering the eight 16D shift vectors $V_A$ that define them, heterotic orbifold models with a given space group $S$ 
exhibit 128 defining parameters. However, as mentioned before, there are highly
nontrivial symmetries relating different gauge embeddings that may reduce
the number of truly free parameters. Noting that the symmetry invariants $N_n^{(a)}$
encode information about the shift vectors, one might entertain the possibility
to replace the original parameters by a set of cardinalities $\{N_n^{(a)}\}$, which
has shown to be a reasonable strategy~\cite{Mutter:2018sra}.

Our purpose here is to design an autoencoder capable of analyzing models of various 
orbifold geometries at once. The challenge here arises from the fact that each
orbifold geometry has very different properties. This implies that the defining
parameters associated with each $S$ exhibit e.g.\ different values and statistical 
characteristics, see our auxiliary material for two sample orbifold geometries~\cite{geometrystatistics}. 
We must find a method to deal with the various geometries simultaneously. For 
instance, since in general the total number of fixed points 
$n_\mathrm{fp}(S)$ depends on $S$, we must restrict the 
selection of fixed points to a subset of size $s_\mathrm{fp}$, such that 
$s_\mathrm{fp} \leq n_\mathrm{fp}(S)$ and is equal for all considered space groups. 
In addition, we must consider a numerical label $L$ for the space group associated with
each model and, in order to capture more features about the 4D spectrum, we shall
include the number of \U1 gauge factors $N_{\U1}$.
Therefore, we choose as characteristic features of a model the subsets
\begin{equation}
\label{eq:featuresX}
X~:=~\left\{N_n^{(a)}~\big|~ 0\leq n\leq s_\mathrm{fp},\ a=1,2\right\} 
    ~\cup~ \Big\{ L, N_{\U1}\Big\}\,.
\end{equation}
Thus, in this scenario, the number of considered features is $2 s_\mathrm{fp}+4$.

There is some arbitrariness on the choice of $s_\mathrm{fp}$ that we can use to our 
advantage. It is clear that the larger $s_\mathrm{fp}$ is, the better $X$ captures
the properties of the orbifold. This is because each chosen fixed point can 
contain information about a single shift vector $V_A$ and we need information about all of them.
However, $s_\mathrm{fp}$ cannot be too large because
there are some space groups $S$ that admit only a small number of fixed points. Further, 
taking the maximal number of fixed points of a given geometry is not necessary in general 
because the definition~\eqref{eq:gaugeEmbeddingTwisted}
of the local gauge embedding implies that there are $V_{\mathrm{local},n}$ at different
singularities that can lead to the same values $N_n^{(a)}$. Moreover, we must also 
take into account that the computation time grows with the dimensionality of $X$. 
So, we find it convenient to consider $s_\mathrm{fp}$ of order 10, and choose the
fixed points from the (first) twisted sectors $(k,\ell)=(1,0)$ and $(0,1)$ in addition 
to some others from higher twisted sectors, in order to capture at various (preferably 
geometrically inequivalent) singularities as much information provided by all shift 
vectors $V_A$ as possible. Note that choosing $s_\mathrm{fp}$ of order 10 
exhausts the information coming from all eight $V_A$ unless we incorrectly take only
geometrically equivalent fixed points.

Once the feature vector $X$ to be used on different orbifold geometries is 
defined, we must build the datasets that will allow us to train the autoencoder.
With this goal, one must first generate a large number (a few hundred thousands)
of inequivalent admissible orbifold models with one or more space groups 
by using the dedicated automatized tool {\tt orbifolder}~\cite{Nilles:2011aj}. This 
generates the ``raw data'' containing a large number of sets of gauge shift embeddings 
$\{V_A\}$. In a second step, we run our \texttt{Makedataset} code~\cite{Makedataset} 
(detailed in section~\ref{sec:Makedataset}) that translates these vectors into 
the sets of gauge momenta $I_n^{(a)}$ defined in eqs.~\eqref{eq:I0} 
and~\eqref{eq:In} and the resulting gauge group in 4D, and then computes the invariant 
feature vectors $X$. 

Due to the differences among the various orbifold geometries, the process 
previously described will, in general, generate very different sets of data for 
different geometries. This is not surprising since the process leading to the cardinalities
$N^{a}_n$ involves information that is particular to the geometry, namely the centralizers and the 
orbifold twisted sectors. Putting all these elements together means that dealing with more 
than one geometry is not just a matter of extending the data to some extra elements of 
a given set, it is a whole new extension to study a different problem. Including more than one geometry gives a more generic perspective of the 4D models arising and, possibly, independent of the geometry. This allows us to explore the heterotic landscape in a somewhat more general perspective.

\subsection{Autoencoder configurations}
\label{sec:GeneralConfiguration}

A deep autoencoder is a feedforward NN, mainly used to dimensionally
reduce or compress some complex high dimensional data. It is built by two components: the 
encoder and the decoder. The purpose of the encoder is to identify and reduce the 
redundancies (and noise) of input data defined by a large number of parameters, lowering 
step-wise in various (hidden) layers the dimensionality of the parameter space. If the encoder is 
deep enough, i.e.\ if it has a large number of layers, and the data is adequate, its 
capability to encode the input data into a small number of parameters tends to improve. 
The last layer of the encoder is known as the (central) {\it latent layer} or latent-space 
representation, and it contains a ``compressed code'' of the input data, represented in 
a small number of parameters. The decoder operates inversely to the encoder, reconstructing 
the data from the latent layer back to its original higher dimensional form.\footnote{Our 
autoencoder is designed to reproduce the input data. However, sometimes autoencoders are 
used to improve the quality of the data by reducing its noise.} One can define the 
accuracy of an autoencoder as the level of likeness between the output resulting
from the decoder and the corresponding original information in the input layer.

Given some input data, one must choose hyperparameters that maximize the accuracy of 
the algorithm. The properties that describe an autoencoder configuration are:
\begin{itemize}
\item Topology: overall structure that defines the way the neurons of the NN
      are connected among different layers. The topology can be symmetric or asymmetric with 
      respect to the latent or bottleneck layer, fully or partly connected, and can include convolutional 
      layers or other types of substructures. We avoid convolutional or other complex layers
      for simplicity.\footnote{Although it might be appealing to implement convolutional 
      layers, we observe that there is no reason to expect ``spatial'' correlation among the 
      parameters as not even shifts and Wilson lines must exhibit correlations to build admissible
      models. As we will see, the use of one-hot encoding, which we need for our categorical 
      features, further justifies our choice of topology.}

\item Architecture: number of layers and number of neurons per layer (layer size). In the case of
      an autoencoder, it includes the sizes of all hidden layers of the encoder and decoder,
      the input and output layers, as well as the size of the latent layer.

\item Initial weight distribution: the values of the trainable parameters or weights that 
      characterize the neurons must be initialized at random values using a method that 
      may be useful to arrive at the best accuracy; it is customary to take a Gaussian or 
      uniform distribution, but other options (such as Xavier or He initializations~\cite{glorot2010understanding,kumar2017weight,koturwar2017weight,datta2020survey,narkhede2022review}) 
      are possible.

\item Activation function: together with a bias, it defines the output of a neuron 
      given some input information; it adds some non-linearity to the learning process in
      order to improve it. Some examples that we shall use in this work include Leaky-ReLU, 
      Softplus, ELU, CELU, ReLU, SELU, Hardtanh, Softmax and LogSigmoid (see e.g.~\cite{clevert2015fast,klambauer2017self,pedamonti2018comparison,hayou2019impact,dubey2021comprehensive} 
      for details of activation functions).
      In principle, every layer can have a different activation function, but we apply, as usual,
      homogeneously the same activation function to all layers for simplicity. Despite the 
      expectation of a poor performance of Hardtanh, Softmax and LogSigmoid, as will be confirmed,
      they can be used here as a basis for comparison.

\item Loss function: evaluation during the training stage that determines the magnitude of
      the inaccuracy that the NN has achieved before updating the weights of the
      network. Motivated by the encouraging results of ref.~\cite{Mutter:2018sra}, where 
      the $L_2$ loss function (equivalent to MSE) provides a good accuracy, we decide to test 
      various loss functions besides the most natural choice of Cross Entropy (CE).
      Some examples of loss functions used in this paper are CE, SmoothL1, 
      MSE, Huber, BCEWL, L1 and Hinge Embedding (see e.g.~\cite{janocha2017loss,nie2018investigation} for details).

\item Optimizer: optimization algorithm used to minimize the loss; some examples applied in this
      work are Adam, AdamW, Adamax, RMSProp, Adagrad, Adadelta, SGD and ASGD (see e.g.\cite{kingma2014adam,wilson2017marginal,choi2019empirical,schneider2019deepobs}
      for details on these functions).

\item Number of epochs: number of times that the algorithm is run to improve the 
      learning skills of the algorithm, trying to minimize the error.

\item Batch size: for each epoch, it is the number of samples in which the training input
      data is split in order to have several training subsets. Typically, large batch sizes
      lead to better statistical characterizations; however, although commonly varied only 
      to adapt it to the availability of memory, we shall see that the choice of batch size 
      can also help to improve the accuracy of the model.

\item Shuffling: to optimize the learning process, whether or not the elements contained 
      in each batch per epoch are randomly shuffled. 

\item Dropout: if applied, it defines the number of dropout layers and the fraction of neurons that are 
      randomly dropped out; this is typically used with the goal of reducing overfitting.
\end{itemize}

The choice of the best configuration can be either ``handcrafted'' by some educated guess or 
achieved by a systematic study of the properties that define the autoencoder. We mix both methods:
we vary systematically three of the configuration parameters (activation and loss functions, and
the optimizer) while fixing all others. In detail, based on the results of~\cite{Mutter:2018sra}, 
we first fix the topology to be feedforward fully connected and symmetric with respect to the latent 
layer, the initial weight distribution to be Gaussian, and we apply shuffling of the data in each epoch.

To fix the architecture, we adapt the feature vectors $X$ so that they can be readily handled by the 
autoencoder.  Since the selected features are all categorical (they admit a limited set of values), 
it is convenient to perform a {\it one-hot encoding} (OHE). To minimize the use of resources as well
as arrive at good results, we take $s_\mathrm{fp}=8$, leading to
18 features which accept various different sets of values $N_n^{(a)}\in[0,240]$. In addition, 
the number of \U1 gauge symmetries in 4D is constrained by $0\leq N_{\U1}\leq16$, and $L$ 
is limited by the number of orbifold geometries that one incorporates in the autoencoder. Each 
feature $X_i$ is then mapped to a subvector $X_{\mathrm{OHE},i}$ of (different) dimensionality $\gamma_i$. 
The resulting full one-hot encoded feature vector $X_\mathrm{OHE}$ has dimensionality $\sum_i \gamma_i =\mathcal{O}(1000)$. 
This is the size of the input and output layers. We shall aim at arriving at a three-dimensional
latent layer, in order to better capture the information from various orbifold geometries. We
shall include 7 or 9 hidden layers, aiming at a NN deep enough to provide admissible results.

Other configuration parameters are defined in various tests, as we now describe.

\subsubsection{Testing configurations}
\label{sec:vanillaTesting}

In order to arrive at the best configuration, we have used orbifold models based on two 
phenomenologically promising and distinct orbifold space groups, which are known as \Z8--I (1,1) and 
\Z{12}--I (1,1) in the notation of ref.~\cite{Fischer:2012qj}. For simplicity, we
label them as \Z8 and \Z{12}, respectively. The features and phenomenology of \Z8 
orbifold compactifications have been studied in detail in~\cite{GrootNibbelink:2013wnc,Olguin-Trejo:2019hxk},
while many interesting properties of \Z{12} models have been presented
in~\cite{Kim:2006hw,Kim:2007mt,Huh:2009nh,Kim:2011yz,Kim:2015mpa,Kim:2017ylg,Jeong:2019zlr}.
It is known that these space groups correspond to two of the most fertile \Z{N} orbifold geometries,
as they yield large sets of MSSM-like models~\cite{Nilles:2014owa,Olguin-Trejo:2018wpw,Parr:2019bta}.
We emphasize that, even though our method is trained here for these geometries, it can be 
generalized to any choice of geometries. While implementing our method for larger sets, one 
must consider that the larger the set of geometries and models is, the more computational 
resources are required.

With the help of the {\tt orbifolder}~\cite{Nilles:2011aj}, we have built a database of 
nearly 270,000 inequivalent, anomaly-free models from each chosen geometry, with several 
different gauge groups $G_\mathrm{4D}$. The total of 540,000 models is randomly split 
into a 66.7\% for the training set and a 33.3\% for the validation set.

In this case, we have considered $s_\mathrm{fp}=8$ fixed points per geometry,\footnote{We 
have selected fixed points from the first and third twisted sectors for \Z8 orbifolds, 
and from the first, second and third twisted sectors for \Z{12} models.} which means 
that the feature vectors $X$ of eq.~\eqref{eq:featuresX} are 20-dimensional. Analyzing the total 
of our models, the OHE enhances the feature vectors to 810-dimensional one-hot encoded vectors 
$X_\mathrm{OHE}$. This represents the input layer.

After some quick tests with various libraries, we found optimal the use of \texttt{PyTorch}~\cite{NEURIPS2019_9015} 
and some \texttt{Pandas}~\cite{mckinney-proc-scipy-2010} and \texttt{Scikit-learn}~\cite{scikit-learn} functions, as they
are simple and economic in notation and as effective as other options.

Now we are ready to implement and test our proposal. Since
we wish to test for several activation, loss and optimizer functions, we define a basic
or ``vanilla'' configuration, which is given in table~\ref{tab:vanilla}. 
As anticipated, we use seven hidden layers, including the 3D latent layer.
The dimensions of the layers of our symmetric NN are 
$810,200,26,13,3,13,26,200,810$. We run the training stage for 1,000 epochs,
with no assumption about overfitting or other undesirable properties that we
want to identify in this phase.

\begin{table}[t!]
\begin{center}
\begin{tabular}{| l | l |} 
 \hline
 Configuration parameters     & Chosen values \\
 \hline
 Topology:                    & Fully connected and symmetric \\
 Architecture:                & $810, 200, 26, 13, 3, 13, 26, 200, 810$ \\ 
 Initial weight distribution: & Gaussian \\ 
 Activation function:         & SELU (default value)  \\
 Loss function:               & MSE (default value) \\ 
 Optimizer:                   & Adam (default value)\\ 
 Number of epochs:            & $1,\!000$ \\
 Batch size:                  & $2^5$\\
 Shuffling:                   & True \\
 Dropout:                     & False \\
 \hline
\end{tabular}
\caption{Parameters of our ``vanilla'' configuration. While others are fixed, we 
vary the parameters with the label ``default value'', one at a time, in order to arrive at the 
best autoencoder configuration. E.g.\ we test different optimizers while taking
the default values specified here for the activation and loss functions.
\label{tab:vanilla}}
\end{center}
\end{table}

We test the different functions of our setup as follows: we choose the one 
function in the vanilla configuration that we want to test and leave the other two fixed 
with the ``default values'' given in table~\ref{tab:vanilla}; then, we vary systematically the 
chosen function through the options listed in section~\ref{sec:GeneralConfiguration}, 
training the NN during 1,000 epochs for each resulting configuration. We tested various
batch sizes ranging from $2^3$ to $2^7$, which are within the capabilities of any small or
large computer, and observed that varying the batch size also allowed us to arrive at a better 
accuracy. Finally, we arrived at a batch size of $2^5=32$ to minimize the hardware demands 
while aiming at good results.\footnote{We performed the training of our NN with an Intel Core i7-6700 CPU, 
32 GB DDR4 RAM. The average training time was about 15 hours.}
The default values are considered taking into account their good performance
in NN for similar purposes and their reduced computing time. In order to evaluate 
the performance of each configuration, we examine the NN accuracy and loss.

While assessing the results, we realized that the typical definition of the loss function, 
which compares the full predicted and input feature vectors, exhibits some weakness because 
of the large number of zeroes appearing in the one-hot encoded vectors that are compared. 
We improve this situation by demanding that the loss function compare the input and output 
one-hot encoded vectors by feature. In detail, we implement the redefinition of the loss
\begin{equation}
\label{eq:loss-redef}
    L(X_\mathrm{OHE},\widehat{X}_\mathrm{OHE}) ~=~ 
                     \frac{1}{\text{\#\,batches}}\sum_k 
                     \frac{1}{\text{batch\,size}}\sum_j 
                     \frac{1}{2 s_\mathrm{fp}+4}\sum_i F(X^{kj}_{\mathrm{OHE},i}, \widehat{X}^{kj}_{\mathrm{OHE},i})\,,
\end{equation}
where $i$ runs over all features, $j$ over all elements of a batch, and $k$ over all batches in the training set.
As usual, $X_\mathrm{OHE}$ denotes the input one-hot encoded feature vector, and $\widehat X_\mathrm{OHE}$
the prediction of the autoencoder. The chunk $X^{kj}_{\mathrm{OHE},i}$ associated with the $i$th feature of
the $j$th element of the $k$th batch has dimensionality $\gamma_i$, which is the number of values that the
$i$th categorical feature admits. The loss function $F(\cdot,\cdot)$ is chosen here to be one of our 
testing set: Cross Entropy (CE), SmoothL1, MSE, Huber, BCEWL, L1 and Hinge Embedding.

\begin{figure}[t!]
\centering
\subfloat[Accuracy by activation function]{
\label{accu_activation}
\includegraphics[width=.5\textwidth]{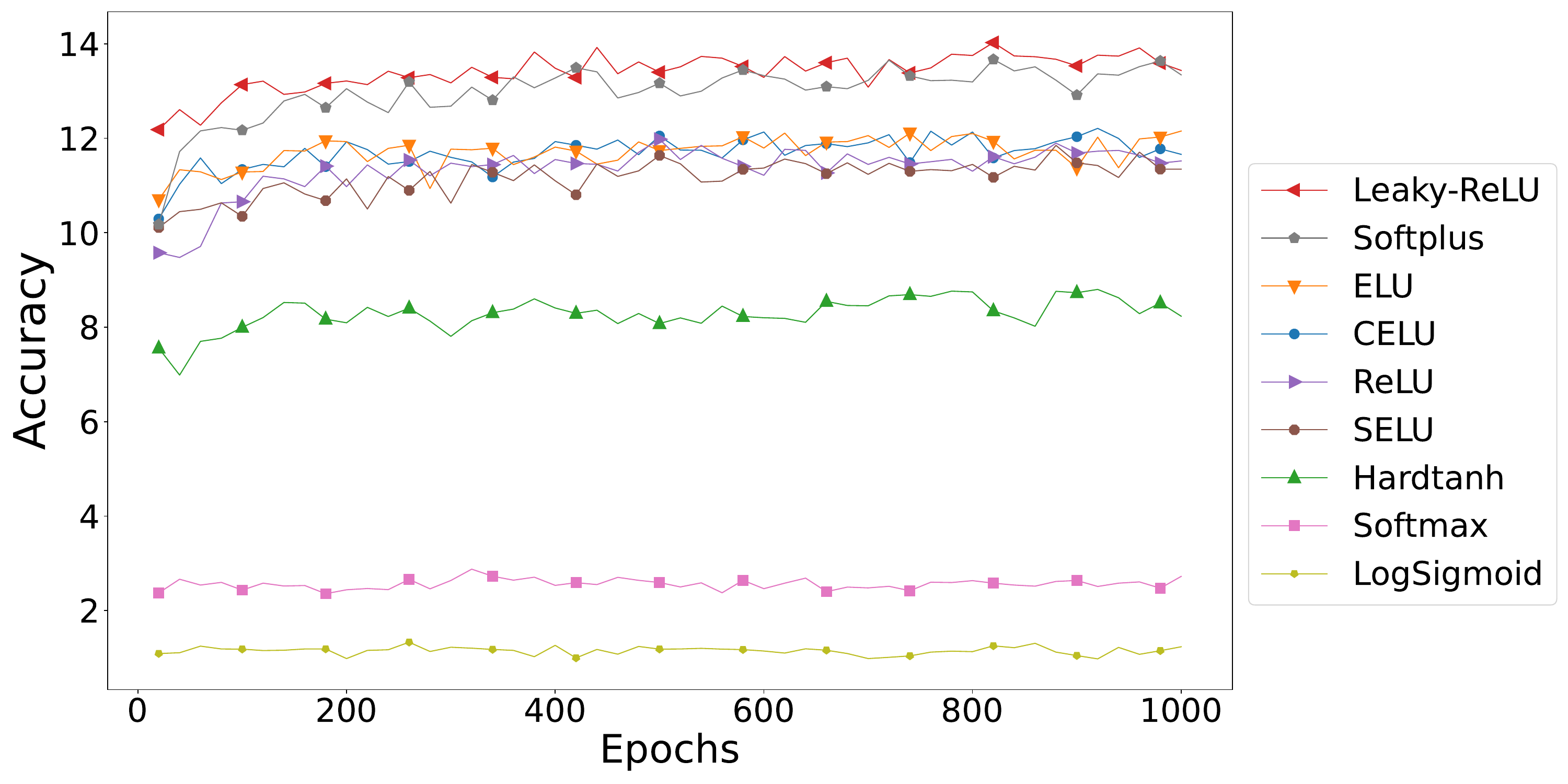}}
\subfloat[Loss by activation function]{
\label{loss_activation}
\includegraphics[width=0.5\textwidth]{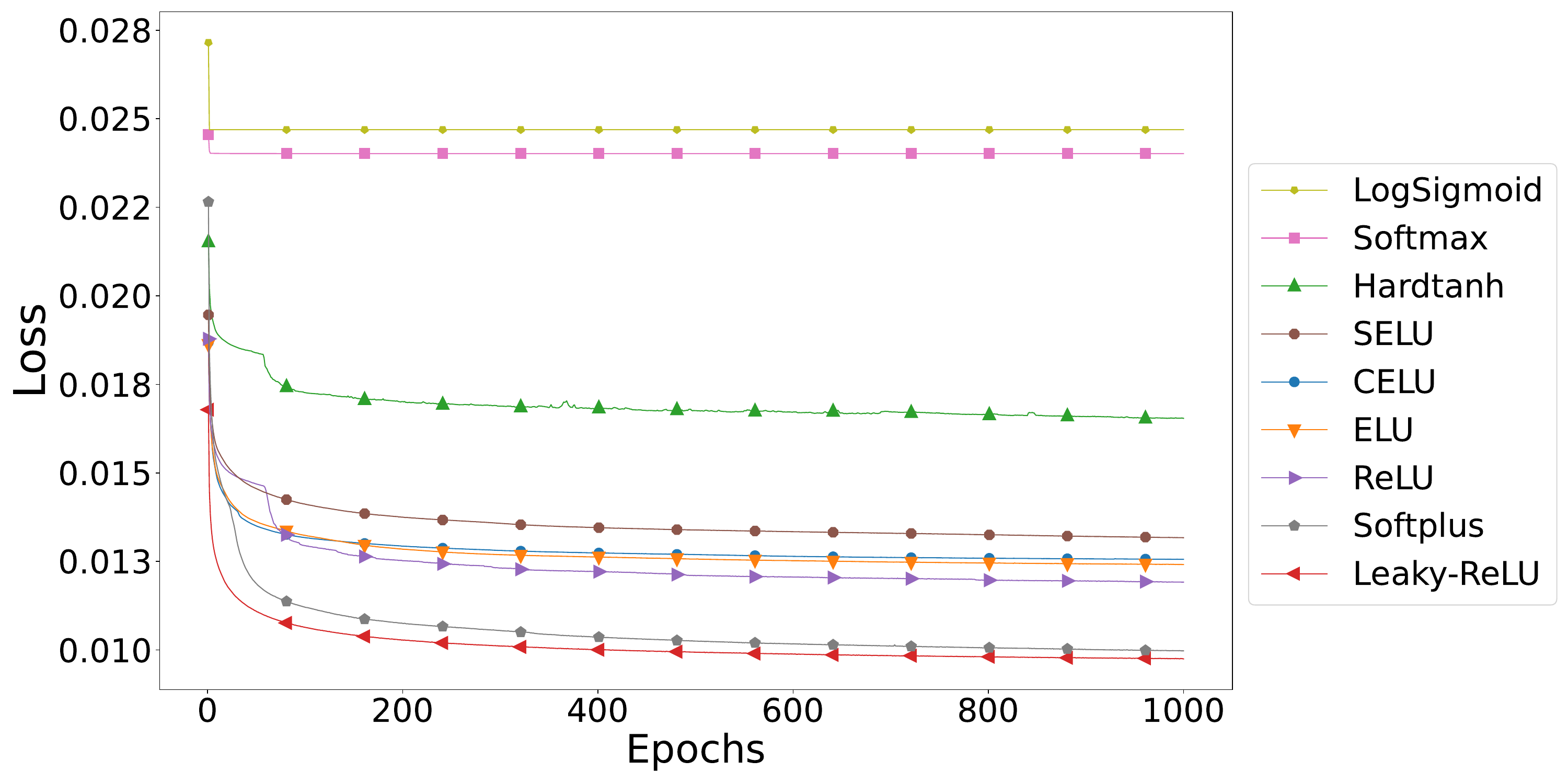}}

\caption{Performance during training of various activation functions on an autoencoder 
based on the vanilla configuration given in table~\ref{tab:vanilla}, with
default values for the loss function and optimizer.
We confirm that the activation functions LogSigmoid, Softmax and Hardtanh
are not efficient for our purposes, as they predict correctly only up to 
eight features out of 20. On the other hand, Softplus and Leaky-ReLU display 
the best performance. These results also allow one to note and compare the 
promising performance of other activation functions when applying deep learning methods.
\label{fig:activations}}
\end{figure}
\begin{figure}[t!]
\centering
\subfloat[Accuracy by loss function]{
\label{accu_loss}
\includegraphics[width=.5\textwidth]{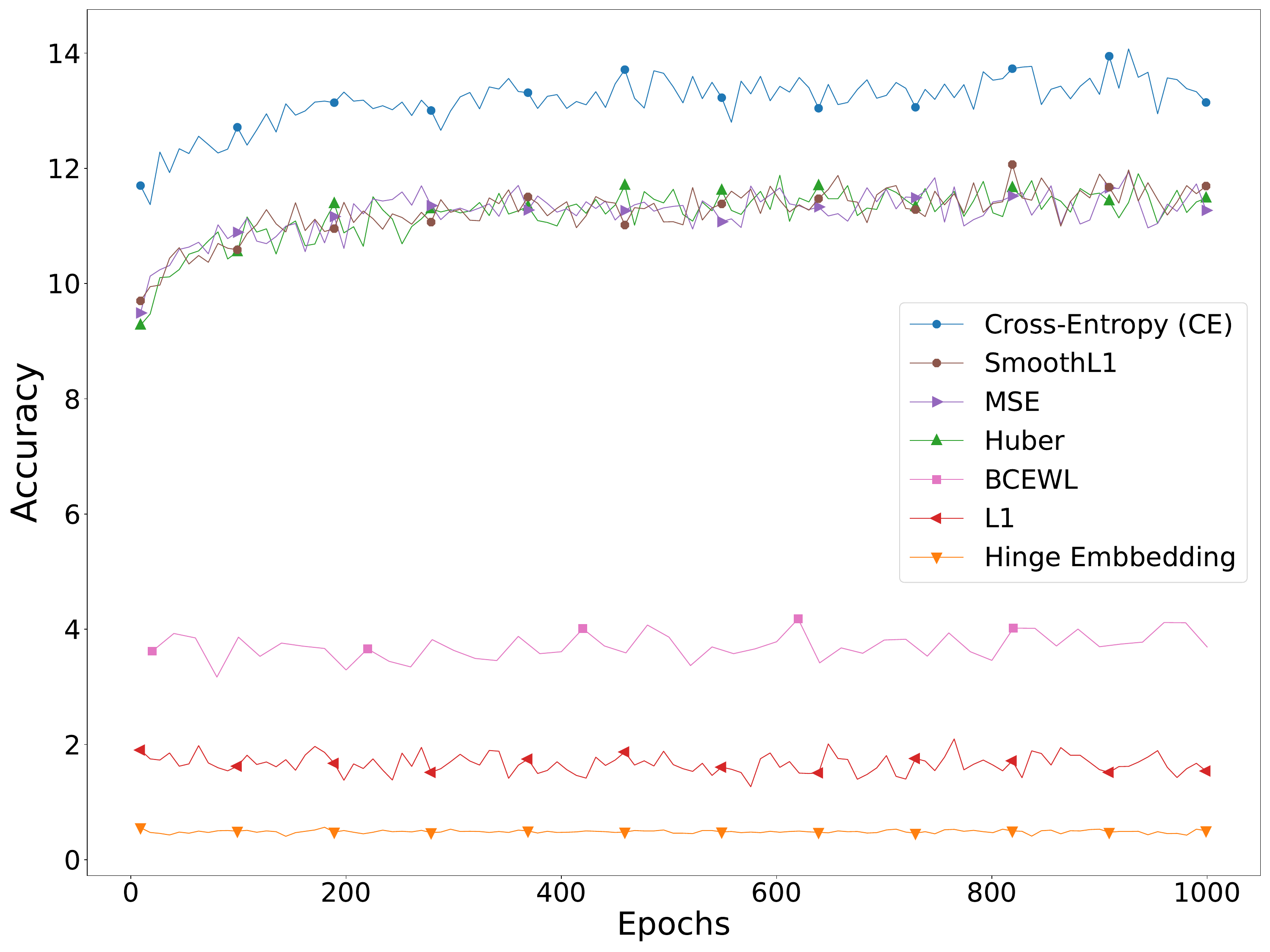} }
\subfloat[Loss by loss function]{
\label{loss_loss}
\includegraphics[width=0.5\textwidth]{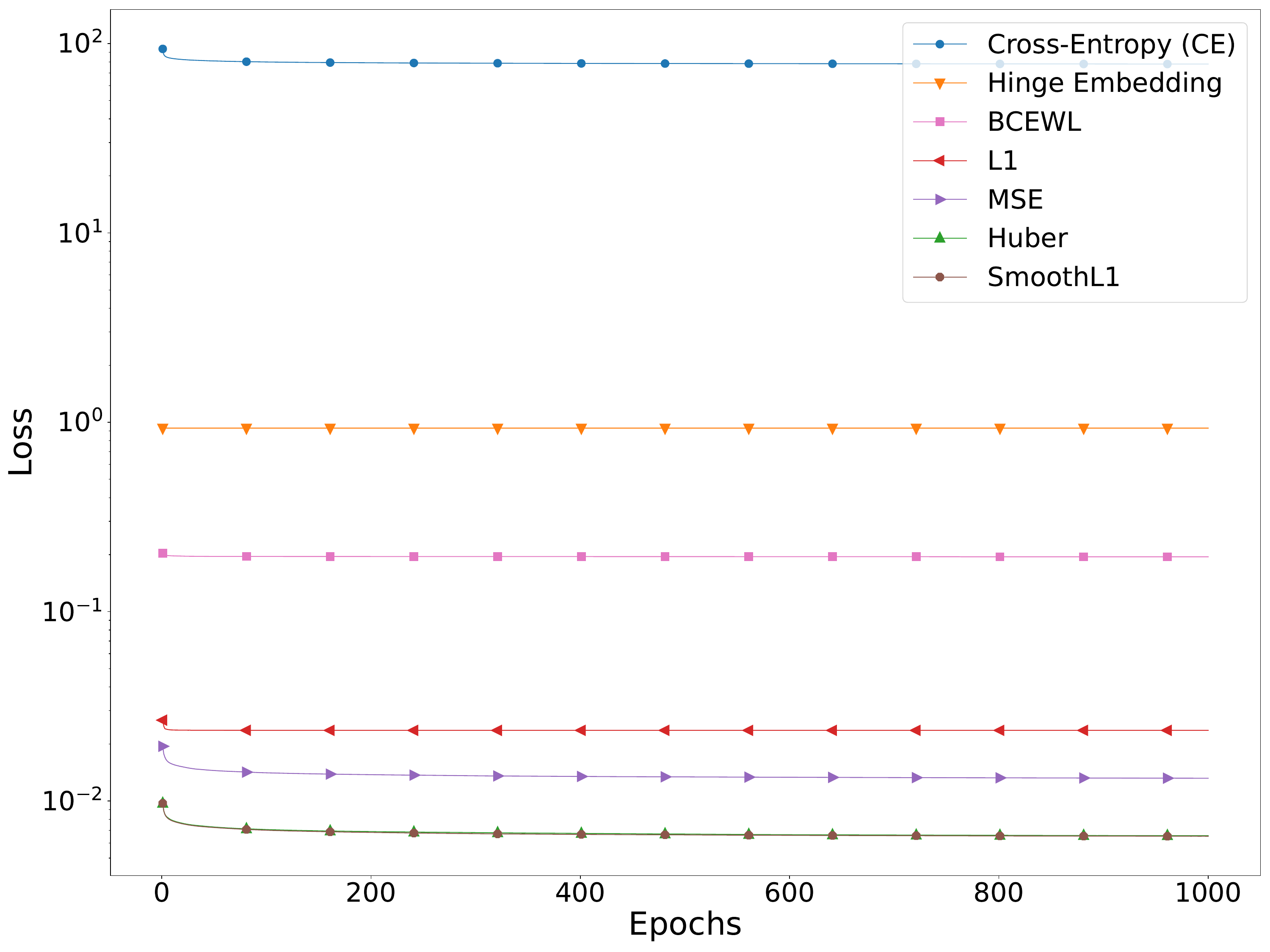}}

\caption{Performance of various loss functions on an autoencoder based 
on the vanilla configuration of table~\ref{tab:vanilla}, with the specified 
default values for the optimizer and activation function.
The L1, Hinge Embedding and BCEWL loss functions reached quickly a
maximal accuracy below $20\%$. On the other hand, CE reached a higher 
accuracy that showed a stable growth, although also the highest 
loss. This must be contrasted with other less accurate with comparable
small loss, such as MSE or SmoothL1. The slope for the various 
loss functions in the loss frame (b) is what matters and not the 
absolute scale.
\label{fig:losses}}
\end{figure}

\begin{figure}[t!]
\centering
\subfloat[Accuracy by optimizer]{
\label{accu_optimizer}
\includegraphics[width=0.45\textwidth]{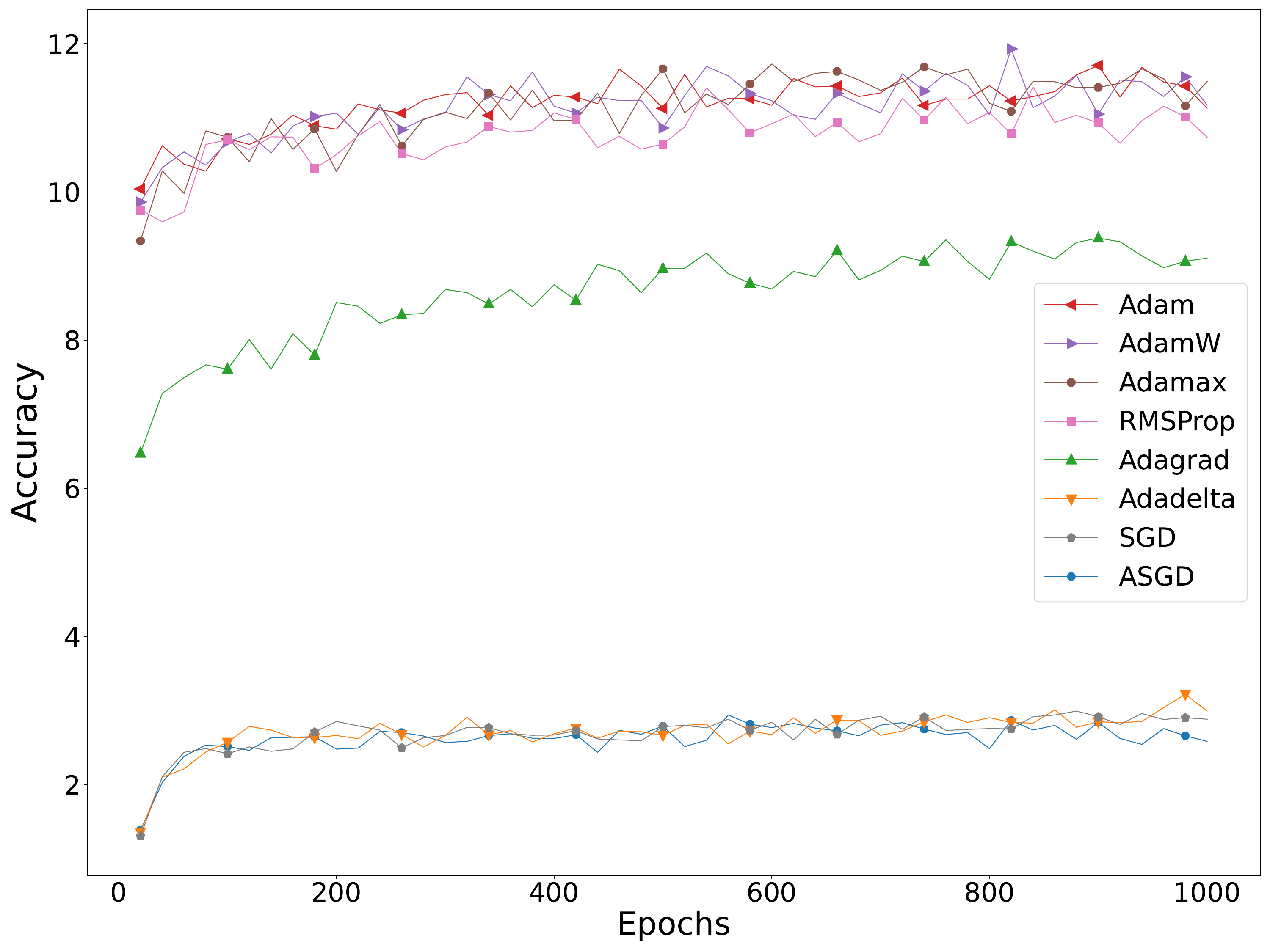}}
\subfloat[Loss by optimizer]{
\label{loss_optimizer}
\includegraphics[width=0.45\textwidth]{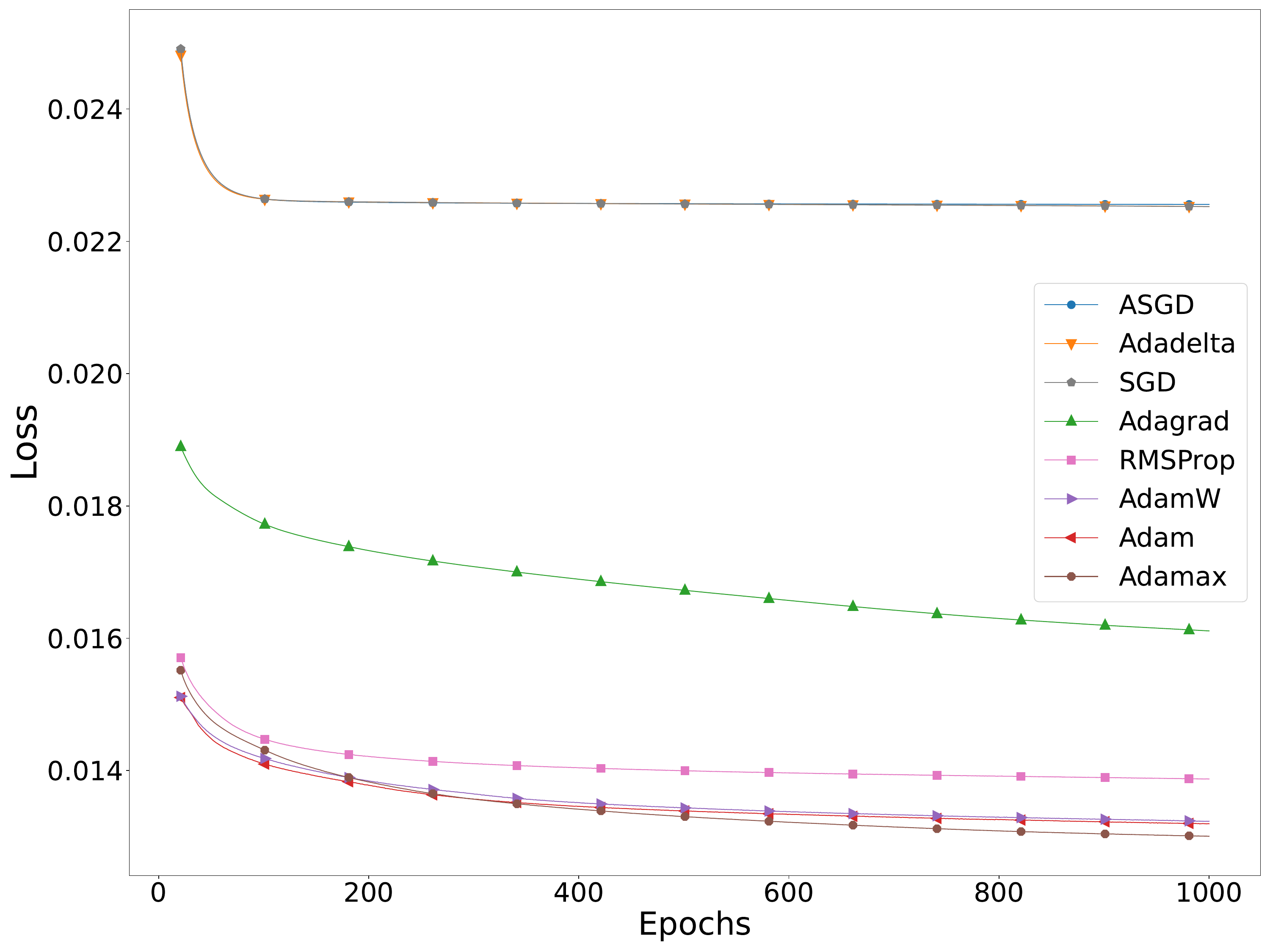}}

\caption{Performance of various optimizers on an autoencoder based on the 
vanilla configuration of table~\ref{tab:vanilla} with default activation and 
loss functions. Adadelta, SGD and ASGD exhibit the worst performance, while
Adam and its variants display the best accuracy and smallest loss. 
}
\label{fig:optimizers}
\end{figure}

Our results at this stage are presented in figures~\ref{fig:activations}--\ref{fig:optimizers}.
In figure~\ref{fig:activations} we show a comparison of the evolution over 1,000 epochs of 
the accuracy and loss of the various activation functions that we consider in our tests. The 
best accuracy is achieved by Softplus and Leaky-ReLU, which corresponds to NN that predict
correctly up to about 14 out of 20 features. Simultaneously, we observe that these activation
functions lower their loss very fast, with a slight preference for Leaky-ReLU, as suggested
in the literature, see e.g.~\cite{pmlr-v15-glorot11a}. It turns out 
that this preference is stressed by the fact that including Softplus in our NN requires much 
longer computation time to arrive at an accuracy comparable with that of Leaky-ReLU.

In figure~\ref{fig:losses}, we compare the accuracy and loss that our autoencoder achieves
for various loss functions along with the default values for activation function and optimizer 
of our vanilla configuration, see table~\ref{tab:vanilla}. The loss has been computed using
the redefinition of eq.~\eqref{eq:loss-redef}. We see that Cross-Entropy (CE) leads to the 
most accurate NN while the loss function keeps decreasing. However, our results suggest 
that it is convenient to explore the performance of other almost as accurate
loss functions. MSE, SmoothL1 and Huber display very similar performance, so that we may 
consider any of them as equivalent for our purposes.

In figure~\ref{fig:optimizers} we present the comparison of the performance of
the autoencoder with various optimizers. We observe that the best accuracy and 
smallest loss, with high variation over the first 400 epochs are Adam, AdamW and 
Adamax. Interestingly though, we note that the maximal accuracy is about 12 out 
of 20 features, which differs from the previous comparisons. This is most likely 
the result of the choice for the default values of loss and activation functions,
see table~\ref{tab:vanilla}. This suggests that we must change our selection
in the final configuration.

\vspace{-2mm}
\subsubsection{Our best autoencoder configuration}
\label{sec:bestConfig}

The results of the previous section do not suffice to identify the best configuration.
We use now a much larger dataset with 1,260,000 \Z8 and \Z{12} orbifold models, split 
as before in a $\nicefrac23$ training and $\nicefrac13$ validation set. We 
consider autoencoder configurations based on our vanilla proposal, table~\ref{tab:vanilla},
with the following modifications:
\begin{itemize}
  \item include a pair of 1620-dimensional hidden layers next to the input and output
        layers as a means to improve the accuracy of the NN~\cite{DBLP:journals/corr/abs-1901-00434};
  \item instead of the ``default values'' of the vanilla configuration, we use the 
        selection of the best activation functions (Leaky-RELU and CELU),
        error functions (CE and MSE), and optimizers (Adam, AdamW and Adamax), 
        and test the performance of the resulting combinations.
\end{itemize}
\vspace{-3mm}
The size of the newly included hidden layers has been tested for best performance.

\begin{figure}[t!]
\centering
\subfloat[Accuracy for the training set]{
\label{AccuracyTrainingCombinations}
\includegraphics[width=0.5\textwidth]{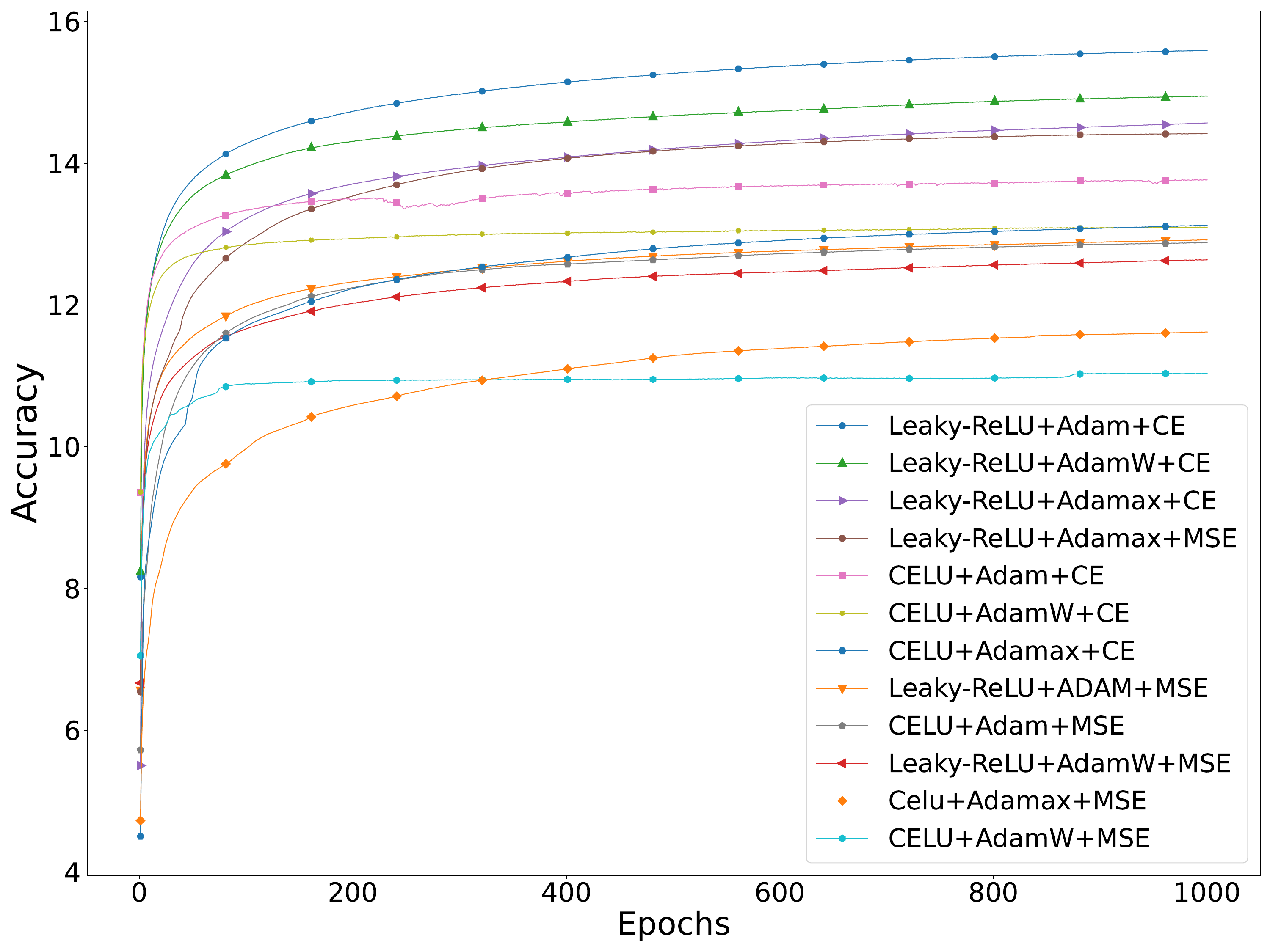}}
\subfloat[Accuracy for the validation set]{
\label{AccuracyValidationCombinations}
\includegraphics[width=0.5\textwidth]{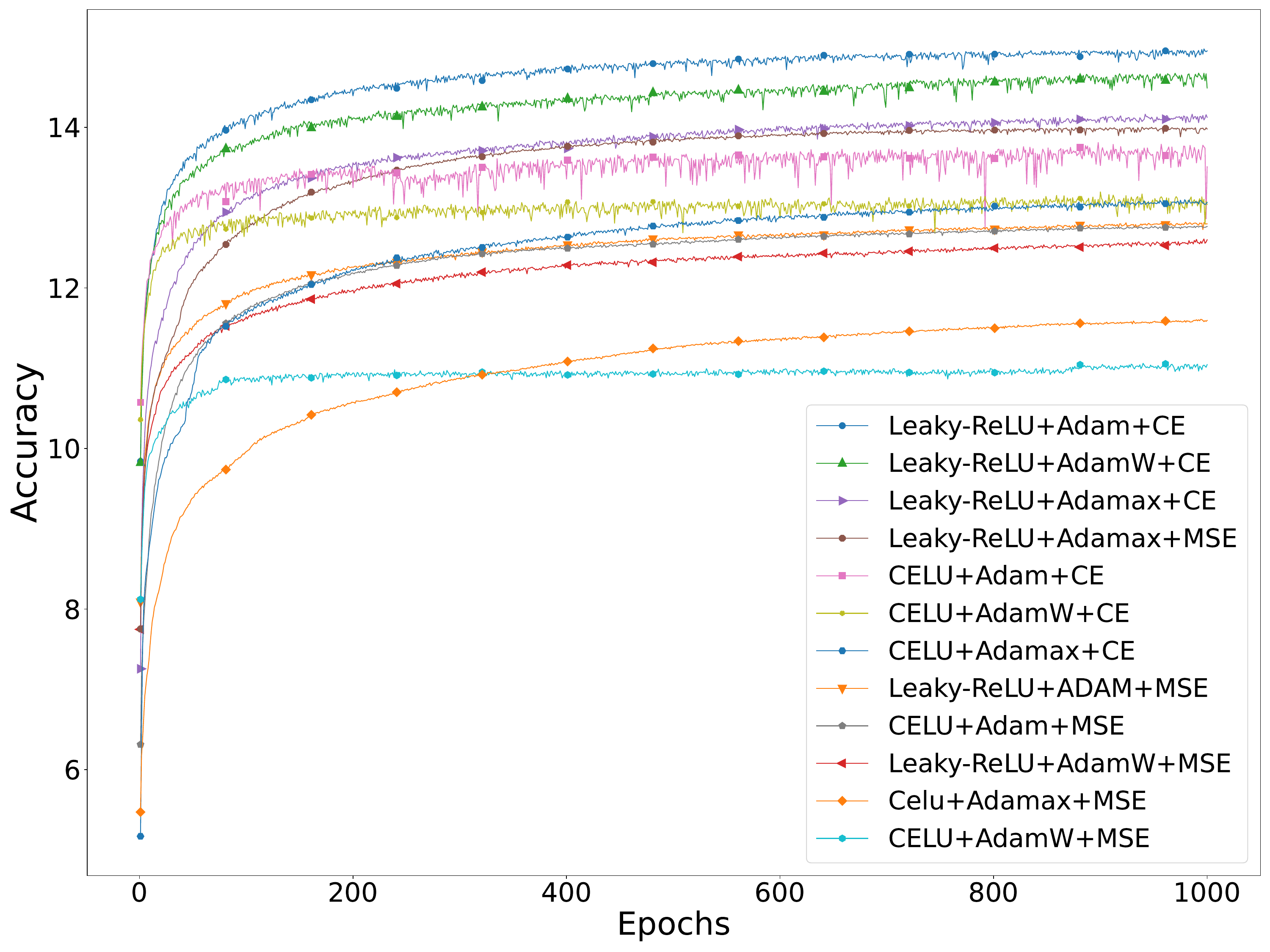}}
\caption{Accuracy of the most promising autoencoder configurations on (a) the training 
set and (b) the validation set. We have included a pair of additional 1620-dimensional 
outermost hidden layers in order to improve accuracy. All configurations perform similarly 
in both sets. The most favored configuration includes Leaky-ReLU universal activation 
function, Adam optimizer and Cross-Entropy error function.
\label{fig:accuracyForCombinations}}
\end{figure}
\begin{figure}[t!]
\centering
\subfloat[Loss for the training set]{
\label{LossTrainingCombinations}
\includegraphics[width=0.5\textwidth]{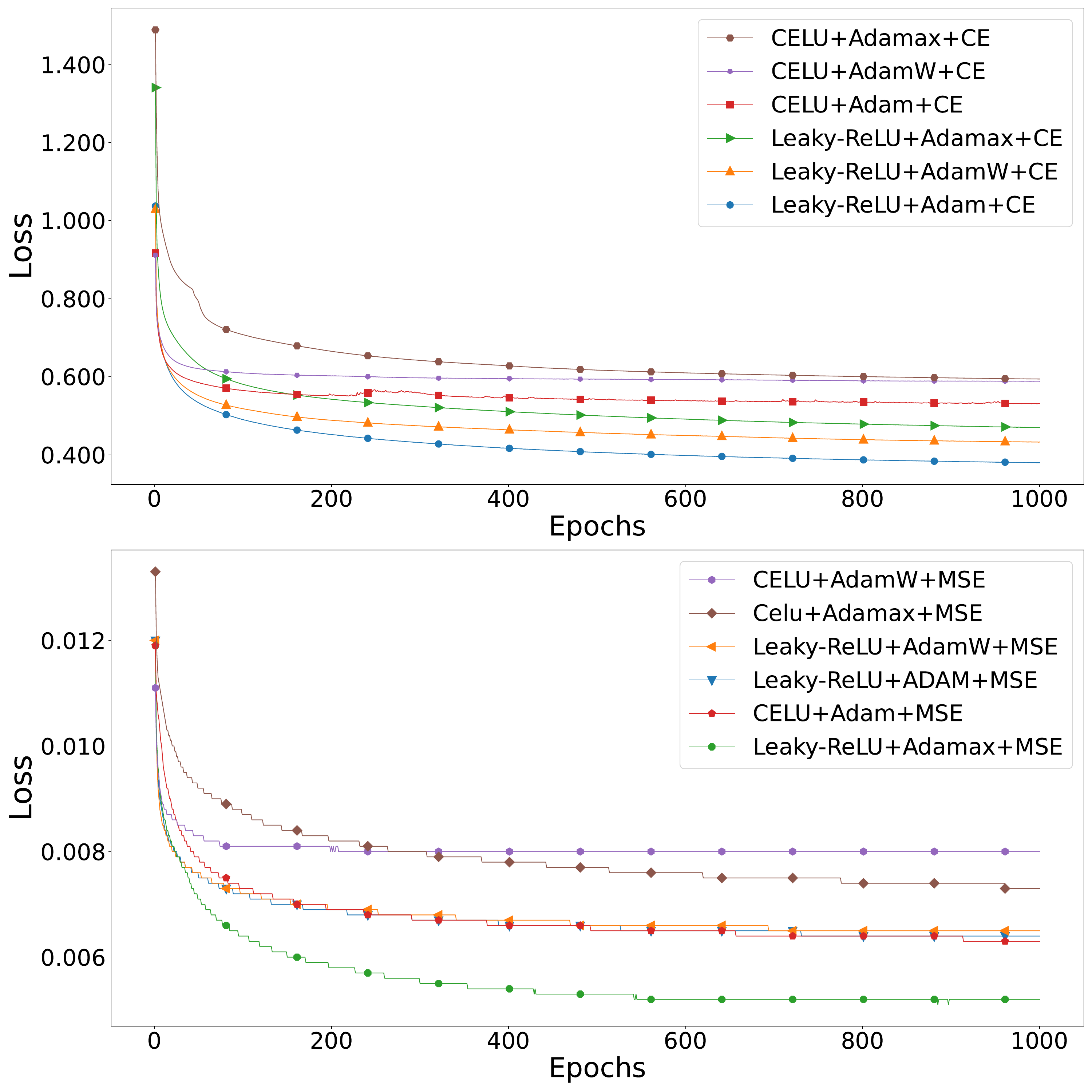}}
\subfloat[Loss for the validation set]{
\label{LossValidationCombinations}
\includegraphics[width=0.5\textwidth]{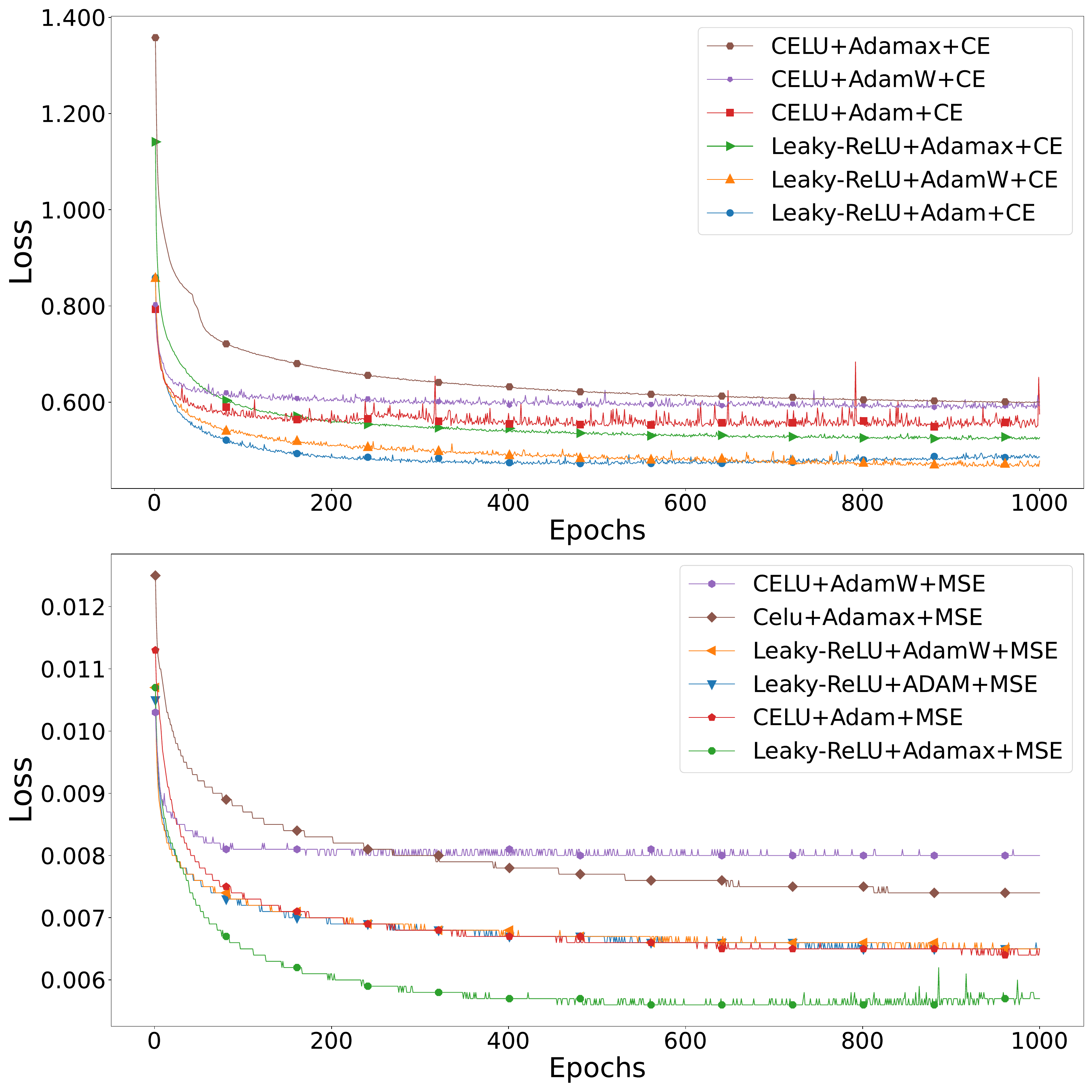}}
\caption{Loss of the most promising autoencoder configurations on (a) the training set 
and (b) the validation set. We split the plots because of the different (arbitrary) scales 
of MSE and CE error functions.
Taking into account the accuracy reported in figure~\ref{fig:accuracyForCombinations},
we see that Leaky-ReLU combined with Adam and Cross Entropy yields the smallest error.
Yet we observe hints of potential overfitting around epoch 400.
\label{fig:lossForCombinations}}
\end{figure}

Given these specifications, we let once more our NN run over 1,000 epochs with each 
of the proposed configurations. In order to better compare the performance, now we
compute the accuracy and loss in both the training set and the validation set.
Our results are displayed in figures~\ref{fig:accuracyForCombinations} 
and~\ref{fig:lossForCombinations}. We first note the stability of our autoencoder
configurations as they all perform similarly in both the training and validation
sets. Further, from the evolution of the accuracy of our NN shown in 
figure~\ref{fig:accuracyForCombinations} we see that the most accurate configurations
always include Leaky-ReLU~\cite{maas2013rectifier} as activation function\footnote{We 
remark that, even though smooth functions are usually favored in deep 
NN~\cite{hayou2019impact, klambauer2017self}, our benchmarking process showed that 
Leaky-ReLU is the most efficient activation function for this particular task, leaving 
the exponential-linear-unit family of functions below by at least 5\% in accuracy after 
epoch 200. This seems compatible with previous findings that favor somewhat 
rectifying neurons when sparse data is analyzed~\cite{pmlr-v15-glorot11a}.}
and CE as error function. All variations of Adam optimizer perform very well. 
Including the evaluation of the loss from figure~\ref{fig:lossForCombinations}, we find 
that the combination of activation, error and optimizer functions and the architecture 
presented in table~\ref{tab:finalconfig}, together with the other NN parameters 
of the vanilla configuration of table~\ref{tab:vanilla},
lead to the best autoencoder configuration. Further, this configuration takes
the most optimal computation time.\footnote{The evaluation for each configuration
took up to 30 hours on a computer with an Intel Core i9-10900K CPU, 
32 GB DDR4 RAM, and Nvidia Quadro P620 GPU (2 GB DDR5, 512 CUDA cores).}

\begin{table}
\begin{center}
\begin{tabular}{| l | l |} 
 \hline
 Configuration parameters    & Best values \\
 \hline
 Architecture:        & $810, 1620, 200, 26, 13, 3, 13, 26, 200, 1620, 810$ \\
 Activation function: & Leaky-ReLU  \\ 
 Loss function:       & Cross-Entropy \\ 
 Optimizer:           & Adam \\ 
 \hline
\end{tabular}
\end{center}
\caption{Best parameters for the autoencoder configuration. Based on the 
various performance tests, we replace the ``default values'' provided in the 
vanilla configuration, table~\ref{tab:vanilla}, by the selection presented
here.
\label{tab:finalconfig}}
\end{table}

\section{Handbook of the \texttt{heterotic orbiencoder}}
\label{sec:HOE}

The main goal of the present work is to introduce a machine capable of 
encoding heterotic orbifold models from various Abelian geometries with the purpose
of analyzing their common properties in a reasonably small parameter space. We release
the generated tools in a couple of easy-to-use and customizable jupyter notebooks: 
\texttt{MakeDataset} and the actual \texttt{heterotic orbiencoder}. The process
is depicted in the flowchart shown in Appendix~\ref{app:data_flow_chart}.

\subsection[MakeDataset]{\texttt{MakeDataset}}
\label{sec:Makedataset}

The first tool that will be required is \texttt{MakeDataset}, which is available 
as a jupyter notebook at~\cite{Makedataset}. It allows one to translate the input shift 
and Wilson lines of orbifold models into feature vectors $X$ to be used by the \texttt{heterotic orbiencoder} 
(see eq.~\eqref{eq:featuresX}), which is a mandatory step prior to train our NN. In
its containing folder, you can find a README file with instructions on how to install 
it. Once installed, you have just to choose five parameters to start the process:
\begin{itemize}
\item Use \texttt{name\_models\_from\_orbifolder}
to set the path and name of the input file that contains all shifts and Wilson lines 
of a given orbifold geometry that will be translated into the characteristic features to
train the \texttt{heterotic orbiencoder}. For example \begin{lstlisting}[style=pyt,numbers=none]
name_models_from_orbifolder = "Z6_models_SP_3.txt"
\end{lstlisting}

\item Use \texttt{name\_geometry} to set the path and name of the orbifold-geometry file 
that corresponds to the models to be read by the \texttt{heterotic orbiencoder}. 
Please, observe that this file must be provided in the precise format defined in the 
\texttt{orbifolder}. For example 
\begin{lstlisting}[style=pyt,numbers=none]
name_geometry = "./Geometry/Geometry_Z6-I_2_1.txt"
\end{lstlisting}
 
\item Use \texttt{name\_dataset} to set the name of the output file. For example 
\begin{lstlisting}[style=pyt,numbers=none]
name_dataset = "Dataset_Z6_I_2_1_[1,3,5].csv" 
\end{lstlisting}

\item Use \texttt{list\_sectors} to set the twisted sectors of the orbifold geometry 
that will be used to generate the characteristic features. It is important to match the 
number of fixed points of the various geometries used. Take into account that the 
number of fixed points depends on the selected sectors. (You can verify it with the command 
\texttt{sector()}, which returns the list of sectors and fixed point for each geometry.)
For example 
\begin{lstlisting}[style=pyt,numbers=none]
list_sectors= [1,3,5] 
\end{lstlisting}

\item Once the process starts, \texttt{MakeDataset} displays the number of feature vectors 
that have been computed and stored. May you need to interrupt this process, you can restart the
generation of feature vectors by restarting the program with an advanced starting point. This is 
set by the parameter \texttt{start}. For example,
\begin{lstlisting}[style=pyt,numbers=none]
start = 100 
\end{lstlisting}
restarts the generation of feature vectors at the 100th model.
\end{itemize}

It is important to mention that one needs first to run \texttt{MakeDataset} for each one of 
the orbifold geometries of interest, then label properly the datasets, and finally concatenate
them. (This can be easily done in \texttt{Pandas}.) The name of the resulting 
concatenated file must feed the variable \texttt{datasetname} of the configuration of the 
\texttt{heterotic orbiencoder}.

\subsection[Heterotic orbiencoder]{\texttt{Heterotic orbiencoder}}
\label{sec:orbiencoder}

The notebook of the \texttt{heterotic orbiencoder} is configured to work directly by loading 
the pre-trained example (Trained\_Machine.pt) used in the present paper. Following the 
standard ML nomenclature, hereafter we will refer to the NN configuration together with a set 
of hyperparameter values as {\it NN-model}. Hence, changing some parameters and then attempting 
to run the default NN-model will unavoidably lead to spurious results.

According to our best configuration given in table~\ref{tab:finalconfig}, the default value for 
the activation function of all layers is set to Leaky-ReLU, and the dimension of the layers, the loss 
function and the optimizer must be fixed as follows

\begin{lstlisting}[style=pyt,numbers=none]
dimensions = [l_ohe, 2*l_ohe, 200, 26, 13, latent, 13, 26, 200, 2*l_ohe, l_ohe]
latent = 3  
criterion = nn.CrossEntropyLoss() 
optimizer = optim.Adam(NN-model.parameters(), lr=1e-4)
\end{lstlisting}
where \texttt{l\_ohe} is the dimension of the one-hot encoded vectors 
$X_\mathrm{OHE}$, obtained automatically when the file \texttt{datasetname} is loaded; 
and \texttt{lr} denotes the optimizer's learning rate, which is
chosen to be $10^{-4}$ to optimize the loss function, based on the previous results.

In addition, we provide a number of training parameters that can be dialed, although
they are set to the best values we identified:
\begin{itemize}
\item Use \texttt{epochs} to set the number of epochs that the NN will be trained. For example 
\begin{lstlisting}[style=pyt,numbers=none]
epochs = 1000 
\end{lstlisting}
\item If you would like to save the resulting NN-model after a number of training epochs, set this 
periodicity in the parameter \texttt{save\_each\_epoch}. For example, to save the NN-model 
every 100 epochs, set
\begin{lstlisting}[style=pyt,numbers=none]
save_each_epoch = 100 
\end{lstlisting}
As an additional feature, the best NN-model, i.e.\ the one leading to the maximal accuracy, is
automatically saved at the end of every training.
\item Use \texttt{train\_set} to set the fraction of the total of the input feature vectors
to be used for training. For example
\begin{lstlisting}[style=pyt,numbers=none]
train_set = 0.6  
\end{lstlisting}
\item Use \texttt{batchsize} to set the size of the batches (number of feature vectors per batch) 
to be used for the training. For example
\begin{lstlisting}[style=pyt,numbers=none]
batchsize = 32 
\end{lstlisting}
\item Use \texttt{label} to set a label for all the output files generated. For example, you may want to
indicate the number of epochs and the activation function to label the files of your training:
\begin{lstlisting}[style=pyt,numbers=none]
label = "1000e_leaky" 
\end{lstlisting}
\item You can use a \texttt{scheduler} to adjust the learning rate after a specified number 
of epochs. This modification in the learning rate can help prevent getting trapped in a local 
minimum during the gradient-descent optimization procedure. A possible configuration of
a \texttt{scheduler} is given by
\begin{lstlisting}[style=pyt, numbers=none]
scheduler = lr_scheduler.StepLR(optimizer, step_size= 400, gamma=0.1) 
\end{lstlisting}
where \texttt{step\_size} sets the periodicity (in the number of epochs) of the change in 
the learning rate, \texttt{gamma} sets the decay of the learning rate, and \texttt{optimizer} 
calls the optimization routine that will be changed every \texttt{step\_size} epochs.
\end{itemize} 
These parameters, along with some others further detailed in the jupyter notebook, 
are contained in a \texttt{Python} dictionary called \texttt{parameters}.

Once the training is finished, you can perform a series of tasks devoted to use the trained
NN-models. First, you can visualize the files of the saved NN-models with 
\begin{lstlisting}[style=pyt,numbers=none]
!ls savedModels
\end{lstlisting}
The NN-model leading to the best accuracy is denoted by ``best'' whereas other NN-models in
the list are displayed according to the epoch number at which they are saved. 

One of the relevant properties of the saved NN-models is their accuracy in 
reconstructing the feature vectors $X$. As expected, every NN-model will have an error. 
To determine it, one can execute the command \texttt{routines.reconstruction}, which
generates two separate files stored in the {\it latentSpace} and {\it success} folders 
within the \texttt{heterotic orbiencoder} path. The first file contains the encoded
feature vectores $X$ in the 3D latent space, while the success file counts the number 
of correctly predicted individual features ($X_i$). 
The command is invoked as follows    
\begin{lstlisting}[style=pyt,numbers=none]
routines.reconstruction(NN-model, saved_model_file_name, **parameters) 
\end{lstlisting}
where \texttt{parameters} refers to the \texttt{Python} dictionary defined in the 
notebook, \texttt{saved\_model\_file\_name} must be taken from the list generated by 
\texttt{!ls savedModels} and \texttt{NN-model} refers to the variable used when the 
NN-model was saved. You can visualize the resulting files by using the commands
\begin{lstlisting}[style=pyt,numbers=none]
!ls success
!ls latentSpaces
\end{lstlisting} 
We also include some functions to visualize useful information regarding the success 
rate and the latent space produced by the saved NN-models, using the files contained in the 
{\it success} and {\it latentSpaces} folders. The commands to invoke those functions are:
\begin{itemize}
\item \texttt{utils.plot\_features\_statistics(success\_file\_name)}: generates bar charts 
with the success rates by feature and the success rates by number of correctly predicted 
features. You can get the \texttt{success\_file\_name} from the list generated by the command 
\texttt{ls sucess}. For example,
\begin{lstlisting}[style=pyt,numbers=none]
utils.plot_features_statistics("success_model-700epoch_1010_e_leaky")
\end{lstlisting}

\item \texttt{utils.plot\_report(label)}: generates the plots of the evolution by epoch 
of accuracy and loss for the validation and training sets. For example, 
\begin{lstlisting}[style=pyt,numbers=none]
utils.plot_report("1000_e_leaky")
\end{lstlisting}

\item \texttt{utils.plot\_2d\_latent\_space(latent\_geom1\_file\_name, latent\_geom2\_file\_name)}: 
generates the latent space plots for the pair of geometries that were fed in to the \texttt{heterotic orbiencoder}. 
You can get \texttt{latent\_geom1\_file\_name} and \texttt{latent\_geom2\_file\_name} from the list 
generated by the command \texttt{ls latentSpaces}.
For example,
\begin{lstlisting}[style=pyt,numbers=none]
utils.plot_2d_latent_space("latent_z8_model-100epoch_1010_e_leaky", "latent_z12_model-100epoch_1010_e_leaky")
\end{lstlisting}
\end{itemize}

An intriguing question arises once the \texttt{heterotic orbiencoder} is trained: Can 
the latent space be used to learn properties of known orbifold models, which were not used 
for training? Can we gain some insight about the properties in the latent space of orbifold 
models with promising properties?
This is particularly relevant when you have a trained NN-model and want to locate MSSM models 
in the latent space that were not included in the training dataset. In this scenario, you can 
copy to the data folder the associated input files containing the feature vectors of the 
promising orbifold models. Without modifying the training parameters, you can select one of 
the saved NN-models and process the new data, granting quick access to relevant reports. 
The parameters and commands to evaluate the parametrical details of the newly inserted
models in the latent spaces are:
\begin{itemize}
\item \texttt{datasetname\_other}: specifies the name of the file that contains the 
promising models that will be encoded into the 3D latent space. For example
\begin{lstlisting}[style=pyt,numbers=none]
datasetname_other = "./data/Z12_SM_554models.csv" 
\end{lstlisting}
\item \texttt{routines.reconstruction\_other\_models(NN-model, saved\_model\_file\_name, datasetname\_other, new\_label, **parameters)}: 
This command performs the same task as \texttt{routines.reconstruction}, but working 
on models unknown during training. The parameter \texttt{new label} sets the name of 
the file containing the data that encodes in the 3D latent space the promising models.
For example
\begin{lstlisting}[style=pyt,numbers=none]
routines.reconstruction_other_models(NN-model, "model-700epoch_1010_e_leaky", datasetname_other,"z12_MSSM", **parameters)
\end{lstlisting}

\item One can generate a plot of the latent space containing the pair of analyzed 
geometries, and additionally the promising models from each geometry analyzed. For this
to work properly, it is important that the labels of the plots be set to MSSM. The 
parameters needed in this function are: \texttt{latent\_geom1\_file\_name}, \texttt{latent\_geom2\_file\_name}, 
\texttt{latent\_MSSM\_geom1\_file\_name} and \texttt{latent\_MSSM\_geom2\_file\_name}, in that specific order. 
For example
\begin{lstlisting}[style=pyt,numbers=none]
utils.plot_2d_latent_mssm("latent_z8_model-100epoch_1010_e_leaky", "latent_z12_model-100epoch_1010_e_leaky", "latent_z8_MSSM-model-100epoch_1010_e_leaky", "latent_z12_MSSM-model-100epoch_1010_e_leaky")
\end{lstlisting}
\end{itemize}

Although we do not recommend to make changes in the configurations, our routines allow 
the user to test different configurations easily. One of the most important characteristics 
to consider is the topology of the \texttt{heterotic orbiencoder}, which is defined 
in the array \texttt{dimensions} as well as the latent space size, defined by the parameter 
\texttt{latent}. Note that changing the size of the latent space will need a consecuent change in the 
functions defined in the module \texttt{utils.py}.

\section{Two orbifold geometries in the eyes of the \texttt{heterotic orbiencoder}}
\label{sec:TwoGeometries}

As anticipated, our autoencoder, based on the configuration of table~\ref{tab:vanilla}
and amended by the improved parameters of table~\ref{tab:finalconfig}, can adopt any number 
of orbifold geometries. For each set of geometries chosen, it must be trained as
we discuss here. The goal of this section is to demonstrate the power of our 
\texttt{heterotic orbiencoder}, while improving over previous results and gaining some physical insight
from the compressed form of the parameter space provided by the autoencoder to 
represent the information of heterotic orbifolds.

Our \texttt{heterotic orbiencoder} is accessible and can be easily executed for any set of orbifold models.
The complete \texttt{Python} code for training the \texttt{heterotic orbiencoder} 
for an arbitrary set of orbifold geometries is made public at~\cite{AutoencoderST}.

\subsection{Final accuracy}
\label{sec:finalAccuracy}

Considering the total of 1,260,000 \Z8 and \Z{12} orbifold models, we aim now at obtaining the 
best configuration of the \texttt{heterotic orbiencoder} for those particular geometries that 
achieves as fast as possible the maximal possible accuracy while avoiding issues such 
as overfitting. We split once more the dataset into $\nicefrac23$ training and $\nicefrac13$
validation set, and train over a number of epochs. Our first step is to train as long as
possible to arrive at high accuracy and observe the behavior of the \texttt{heterotic orbiencoder}. Yet in 
appendix~\ref{app:overfitting} we show that it is very easy to fall into overfitting. 
As we discuss in our appendix, we have tested various methods to reduce overfitting, 
which turn out to be unsuccessful in our case, implying a positive observation though:
The chosen configuration parameters, specifically those of table~\ref{tab:finalconfig},
allow one to obtain high accuracy in short time, before overfitting sets in. In order
to avoid it while delivering the best results, our autoencoder must be trained for 
less than 448 epochs. We saved the best trained NN-model at epoch 445; you can find it 
in the folder saved models in the documentation~\cite{AutoencoderST}.

\begin{figure}[b!]
\centering
\subfloat[Best training]{
\label{fig:finalTraining}
\includegraphics[width=.58\textwidth]{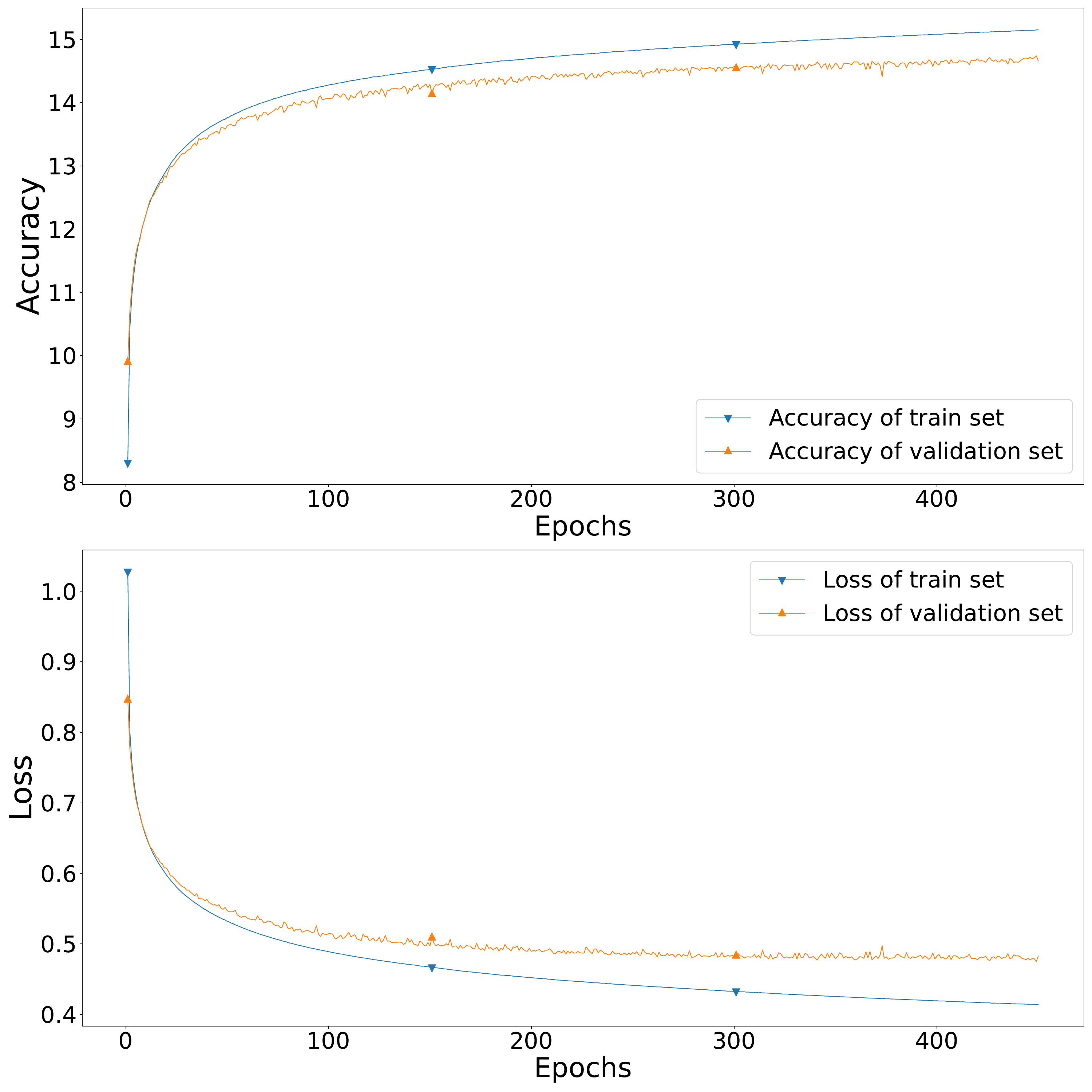}}
\subfloat[Accurately predicted features]{
\label{fig:PercentageAccuratelyPredictedFeatures}
\hspace{4mm}
\includegraphics[width=0.32\textwidth]{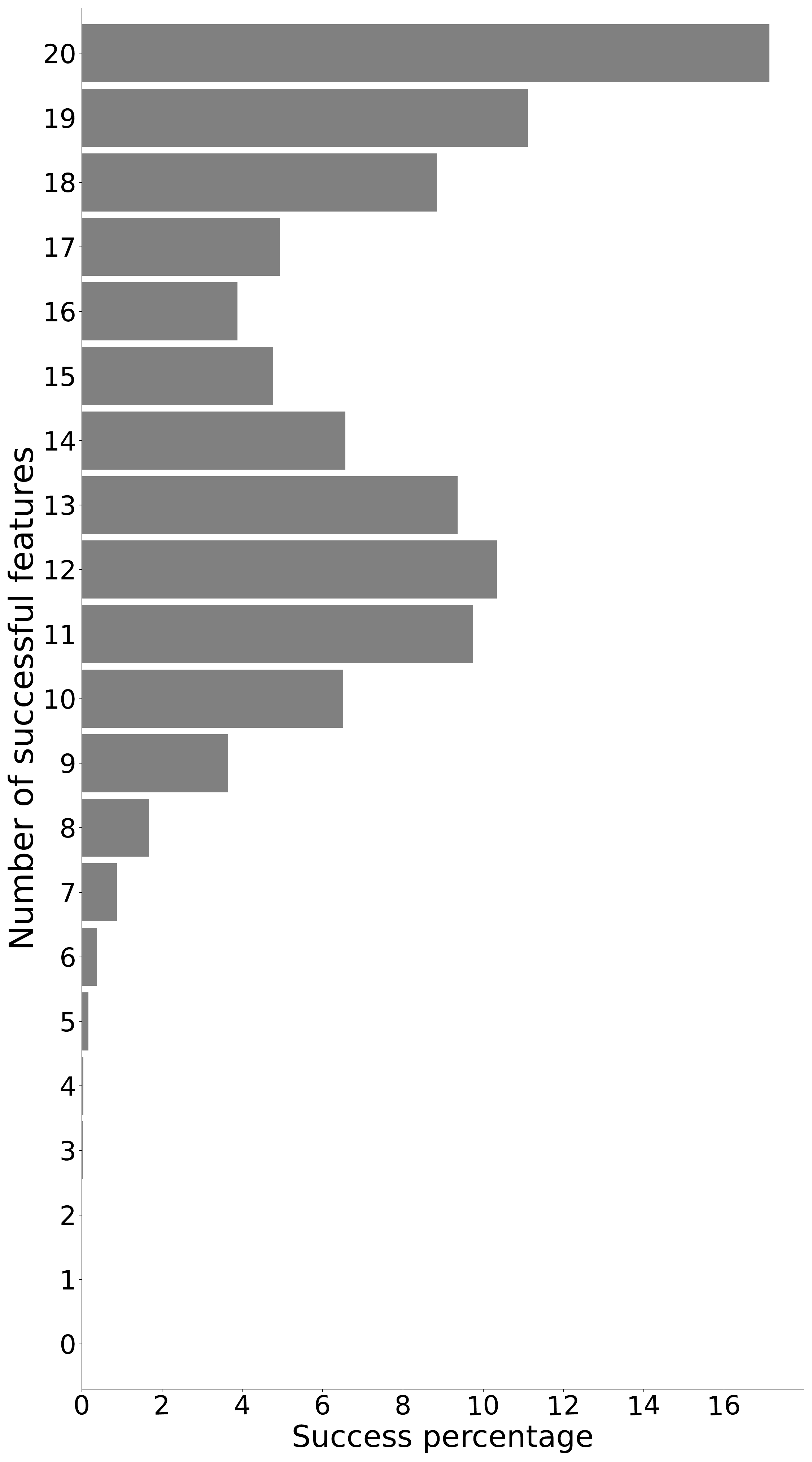}
\hspace{4mm}\phantom.}
\caption{(a) Performance of an orbifold autoencoder trained with the parameters of 
table~\ref{tab:finalconfig}. From our appendix~\ref{app:overfitting} we learn that 
the maximal accuracy that avoids overfitting is reached at epoch 448. Using
an autoencoder trained over 445 epochs, we find that 72\% (75\%) accuracy 
is reached for the validation (training) set.
(b) Frequency of accurately predicted features in all orbifold models. 18 or
more features are correctly reconstructed in 37\% of the models from the 
3D latent space, and at least five features are always correctly predicted.
\label{fig:finalStats}}
\end{figure}

After reducing the $\mathcal{O}(100)$ parameters that define heterotic orbifold models 
to just 20 features of the kind introduced in eq.~\eqref{eq:featuresX}, as before, 
our choice of autoencoder configuration can further encode them in just three parameters,
delivering the final accuracy displayed in figure~\ref{fig:finalStats}.
From a 3D latent layer our decoder can reconstruct in average 14.4 out of 20 features
successfully, corresponding to a 72\% of accuracy. Moreover, we note from 
figure~\ref{fig:PercentageAccuratelyPredictedFeatures} that in 37\% (46\%) 
of \Z8 and \Z{12} orbifold models our autoencoder can successfully reconstruct 
18 (16) or more out of 20 features from the three parameters of the latent layer.
Further, there are models in which only five features are correctly predicted; 
however, they are a very small fraction of the models.

In previous efforts~\cite{Mutter:2018sra}, studying just one orbifold geometry 
with a different NN configuration, an average of 63\% accuracy was achieved 
when using a 2D latent layer. We have made a test to verify whether the 
extra dimension of the latent layer is responsible for the improvement we
see. We found that, in the case of two geometries, considering the activation
function and optimizer of that work and a 2D latent layer leads to a worse
accuracy than the one reported. Further, insisting in a 2D latent layer, but 
keeping other functions as in our configuration leads to a 10\% improvement.
This delivers some interesting observations: our selection of 
configuration parameters is the key to improve accuracy, and one needs more than two parameters
to encode the information of heterotic orbifold models. We must emphasize that our results
are obtained over models from two orbifold geometries (and that this can be readily generalized 
to more geometries with our code), whereas previous results were only on one geometry.

We expect that our results can be improved by making our autoencoder even deeper, 
which will be studied elsewhere by getting access to richer hardware resources 
than currently available to us.

\subsection{Physical insight from the latent space}
\label{sec:latentSpace}

\begin{figure}[t!]
    \begin{center}
     \includegraphics[scale=0.1]{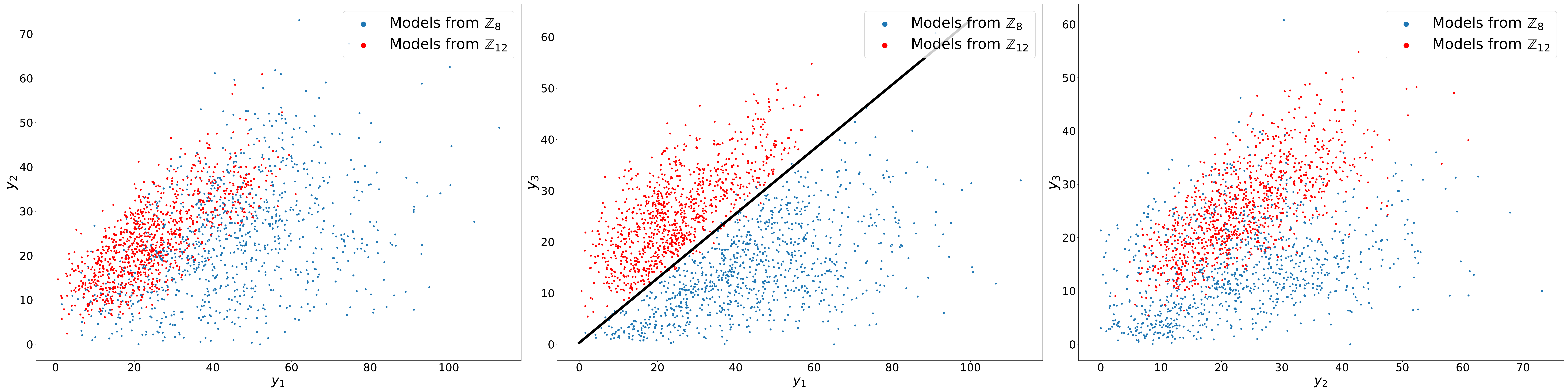}
    \end{center}
    \caption{Three orthogonal 2D projections of the localization of orbifold models in the 
    3D latent space. We present 1,000 sample models of the \Z8 (blue) and \Z{12} (red)
    orbifold geometries. Note that the second projection clearly shows that the 
    geometries separate in latent space. We observe that this segmentation is 
    subtle and would not be direct to identify, should we avoid the label for the \Z8 and 
    \Z{12} models. However, once stablished the separation between different orbifolds,
    some interesting information about the most phenomenologically suitable models will
    arise, as we shall see in figure~\ref{fig:MSSMlatent}.
    \label{fig:generalLatentSpace}}
\end{figure}

In order to learn to draw some information about heterotic orbifold models 
from the three parameters (say $\{y_1,y_2,y_3\}$) of the latent parameter space
of our autoencoder, we must explore the behavior of our models in the latent layer. 
In figure~\ref{fig:generalLatentSpace} we display the localization in the latent 
layer of a sample of 2,000 (randomly chosen) \Z8 and \Z{12} orbifold models per geometry. 
We observe an interesting segmentation of the models by geometry (i.e.~by space group), 
which is evident in the plane $y_1-y_3$. This property most certainly shows that the
information of the feature $X_i=L$ of eq.~\eqref{eq:featuresX} is correctly encoded in
the latent space. The boundary of the two regions is described by the curve given by
\begin{equation}
\label{eq:frontier}
  y_3 - 0.63y_1 = 0.32\,,
\end{equation}
that is also plotted in the central panel of figure~\ref{fig:generalLatentSpace}.

Another interesting observation is that the distribution of the models in the 3D latent 
space is not so wide, so that a classification of heterotic orbifold models would just 
require a scan of a relatively small 3D region, which is much more feasible than the 
same task in the original $\mathcal{O}(100)$-dimensional parameter space. This is the 
motivation of a larger work in progress, which will be reported elsewhere. An important
pending task is the translation of the information contained in the outer layer back to
the parameters that define an orbifold compactification.

Although this information is already important, we would like to explore whether one
can identify some properties of e.g.\ phenomenologically promising models in this new 
parameter space. With this purpose, we use the \texttt{orbifolder} to produce the 
128 input parameters of a set of promising \Z8 and \Z{12} orbifold models with 
the matter spectra of the MSSM or \SU5 GUTs~\cite{Georgi:1974sy}. It is crucial to 
emphasize that these models were not part of the original dataset; hence, they were 
neither included in the training nor validation sets. So, these models can be considered
as a test set.

\subsubsection*{MSSM-like orbifold models}

\begin{figure}[b!]
    \begin{center}
     \includegraphics[scale=0.107]{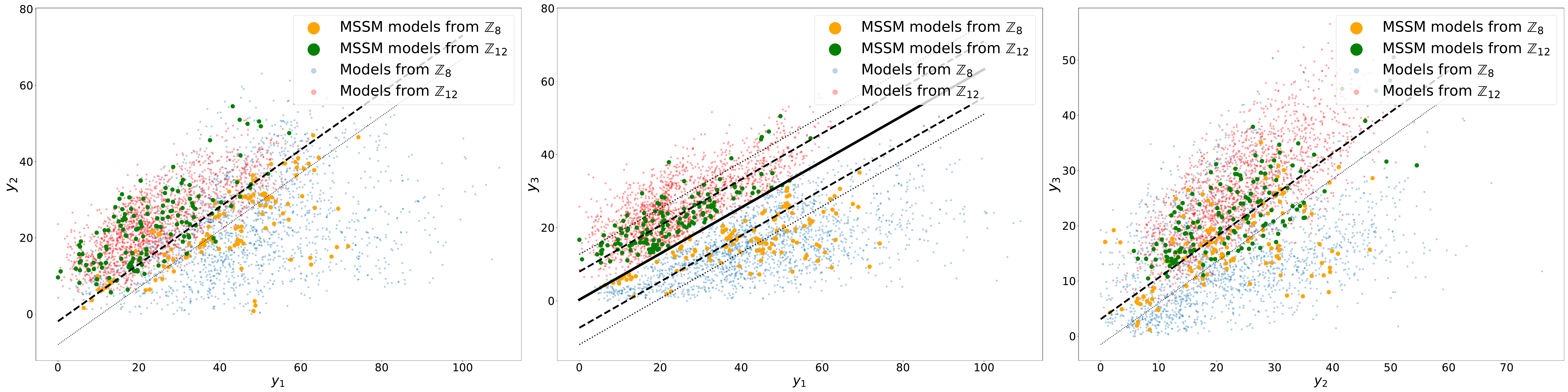}
    \end{center}
    \caption{On top of the localization of arbitrary \Z8 (blue) and \Z{12} (red) 
    heterotic orbifold models, we display where MSSM-like models emerging from the 
    \Z8 (yellow) and \Z{12} (green) orbifold geometry are found in the 3D latent space.} 
    \label{fig:MSSMlatent}
\end{figure}

We have used the \texttt{orbifolder} to construct MSSM-like \Z8 and \Z{12} orbifold
models. (These models had been previously found in ref.~\cite{Olguin-Trejo:2018wpw}.)
We identify 176 (554) models of this type arising from the \Z8 (\Z{12}) orbifold
geometry. In order to present a balanced statistics of the behavior of our models,
we display in figure~\ref{fig:MSSMlatent} the localization in the latent space of 
150 models per geometry.

By observing the details of the localization, we find that close to the boundary
between the \Z8 and \Z{12} orbifold regions, bounded by the line described by
eq.~\eqref{eq:frontier}, most of the MSSM-like models are found. To see this in
more detail, in the plane $y_1-y_3$ (central panel) of figure~\ref{fig:MSSMlatent} 
we depict by the dashed (dotted) lines the boundaries where 50\% (75\%) of all 
promising models are found. In the units of these parameters, the distance from 
the central boundary to the dashed (dotted) lines is 7 (11).
In the $y_1-y_2$ (left panel) and $y_2-y_3$ (right) latent planes, we find that,
even though the boundary between \Z8 and \Z{12} are not distinguishable from
this perspective, on the left of the dashed (dotted) lines 50\% (75\%) of all
MSSM-like models, including both orbifold geometries. This description holds 
for all MSSM-like models from the inspected orbifold geometries.

Our observations imply that a search for promising models is quite restricted 
in the latent layer of our autoencoder. This is unexpected since no information about
the phenomenological features of the models was provided in the input dataset.
Hence, this invites to further explore the question of a better classification
method of heterotic orbifolds, independently of the geometry, through the 
new parameter space of the latent layer.

We note that, in contrast to the results of ref.~\cite{Mutter:2018sra}, no
islands are built in the latent parameter space. We find that this is a general
feature of our autoencoder. It would be interesting to figure out under which
conditions the segmentation behavior arises.

\subsubsection*{\boldmath \SU5 grand unified models \unboldmath}

\begin{figure}[h!]
    \begin{center}
     \includegraphics[scale=0.107]{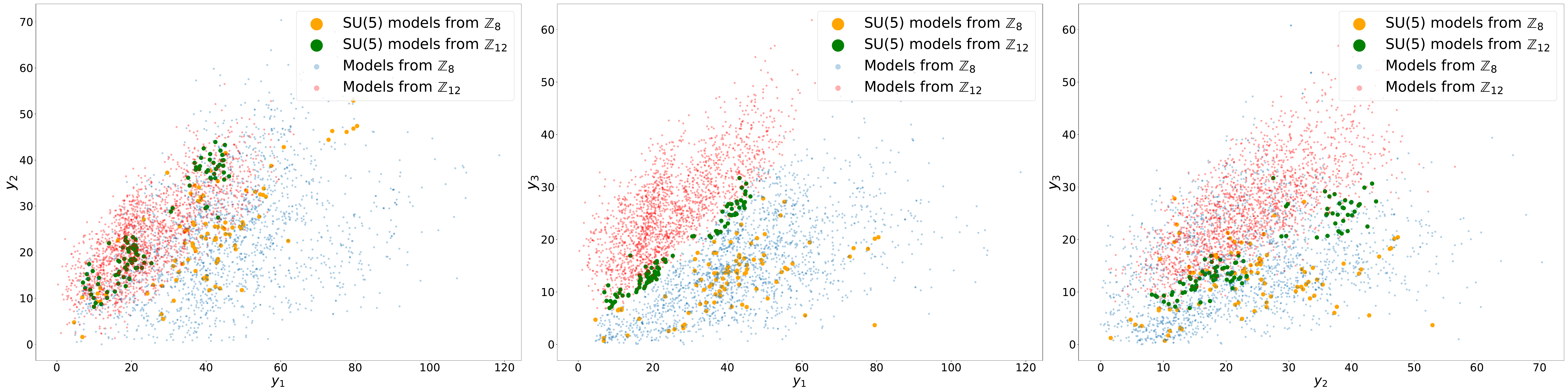}
    \end{center}
    \caption{On top of the localization of arbitrary \Z8 (blue) and \Z{12} (red) 
    heterotic orbifold models, we display where \SU5 GUT models emerging from the 
    \Z8 (yellow) and \Z{12} (green) orbifold geometry are found in the 3D latent space.} 
    \label{fig:SU5latent}
\end{figure}

In figure~\ref{fig:SU5latent} we plot the localization in the latent space of 100 models
per geometry that yield the massless matter spectrum of \SU5 GUTs. As shown explicitly 
in appendix~\ref{sec:SU5} for two sample models, we find that {\it most} of the models of this 
type are collected close to the boundary between the \Z8 and \Z{12} regions, described
by eq.~\eqref{eq:frontier}. This behavior shows that the observations made
for MSSM-like models are repeated for other promising compactifications.

We verified the robustness and replicability of our observations, thus disentangling 
them from a mere stochastic result. We tested different trainings, run on different 
hardware, by varying initialization weights while keeping everything else fixed. All the 
tests showed that, although the latent space had variations, there was always a clear boundary 
between models arising from different geometries and most of the promising models appeared
located near to it. For the curious reader, some of our tests can be found in our 
auxiliary material~\cite{robustnesstests}.
\section{Conclusions and Outlook}
\label{sec:conclusions}

One of the challenges in string model building is the classification
of large classes of consistent models. The reason is the huge dimensionality of the 
input-parameter space where models are specified. For example, heterotic orbifold 
models are defined by $\mathcal{O}(100)$ parameters which can take rational values
subject to some consistency constraints. With the goal of improving this situation,
we have introduced a new general tool based on ML that helps reduce understand these
constructions in a smaller parameter space.

By inspecting several possible autoencoder configurations, we have built 
a robust ML tool known as \texttt{heterotic orbiencoder} (available at~\cite{AutoencoderST}), 
that allows one to encode into just three parameters the information of models arising from an 
arbitrary number of Abelian orbifold geometries, irrespectively of the strong differences
in their geometric and statistical features. Further, it is possible to establish some 
properties of promising models in the smaller latent space. Interestingly, we have shown 
that (with the right choice of activation and loss functions along with the proper optimizer and 
NN architecture) the \texttt{heterotic orbiencoder} does not require long training times 
to render the maximal accuracy of the machine, allowing the training process to be executed 
on an easily accessible computer.

We have applied our \texttt{heterotic orbiencoder} 
to the \Z8--I (1,1) and \Z{12}--I (1,1) orbifold geometries, and
found that this neural network can reconstruct with an average accuracy of 72\% 
the input information from those three parameters. Interestingly, we find that 
in this 3D space (i) models split naturally by geometry, building compact 
clusters, and that (ii) at their borders one can identify phenomenologically 
appealing models, such as those exhibiting the massless matter spectrum of 
the MSSM or \SU5 GUTs. This curious feature reveals that the input parameters
contain already implicit data about the phenomenological properties of the 
models, which is inherited to the 3D latent space. We inspected explicitly 
the translation between the input and latent spaces in \SU5 orbifold GUTs.

Our findings suggest a new classification method based on exploring the 
properties of the latent space. This task seems more feasible than the quest of
obtaining all possible admissible values of the $\mathcal{O}(100)$ original
input parameters. A first short-term goal is to improve the accuracy of
our autoencoder. We expect this to be reached by adding more hidden layers
to our configuration, using larger computing resources than those available 
for the present work. A second goal will be to understand how consistency
conditions, such as modular invariance, are translated into the compressed
information contained in the small latent space. The results of this 
endeavor will be reported elsewhere.

The success of these deep-learning techniques opens up new possibilities. One 
interesting idea is to implement Variational Autoencoders, Generative Adversarial 
Networks, or diffusion models,
that use known sets of string models to produce new unknown models
with selected features. This is the goal of ongoing research that shall be
presented in a coming paper.

\subsection*{Acknowledgments}
It is a pleasure to thank Jos\'e Alberto V\'azquez and Xim Bokhimi for providing 
hardware support for this project. This work was partially supported by 
UNAM-PAPIIT IN113223, CONACYT grant CB-2017-2018/A1-S-13051 and Marcos
Moshinsky Foundation.

\newpage
\appendix

\begin{landscape}
\section{\texttt{Heterotic orbiencoder} flowchart}
\label{app:data_flow_chart}

\begin{figure}[h]
    \centering
    \includegraphics[width=1.32\textwidth, angle=0]{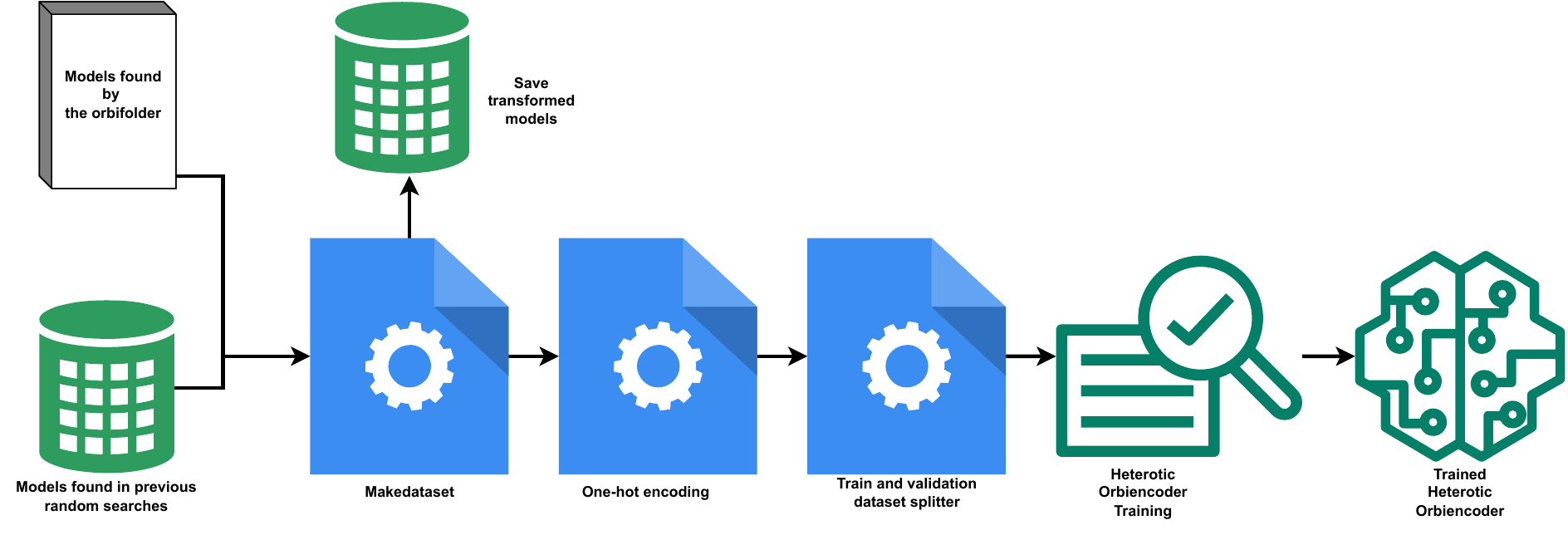} 
    \caption{Description of the process to arrive at the \texttt{heterotic orbiencoder}.
    The input data consists of the 16D shift vectors and Wilson lines that define heterotic 
    orbifold models. These vectors are obtained from previous searches and/or
    the use of the \texttt{orbifolder}~\cite{Nilles:2011aj}. In order to avoid redundancies,
    equivalent models are filtered out while mixing the various models. Subsequently, 
    the orbifold data is processed by our \texttt{Makedataset} code~\cite{Makedataset}
    in order to produce a $\mathcal{O}(20)$-dimensional feature vectors $X$ that comprise 
    the input for the \texttt{heterotic orbiencoder}, as detailed in section~\ref{sec:dataset}. 
    This dataset is stored and one-hot encoded, producing
    $\mathcal{O}(1000)$-dimensional vectors $X_\mathrm{OHE}$. The set of all 
    OHE-processed vectors is split into a large training dataset (we take $\nicefrac23$ of the total)
    and a validation dataset (we take $\nicefrac13$ of the total). The whole data is used as input of 
    the autoencoding NN with the parameters described in section~\ref{sec:bestConfig}.
    By choosing the the hyperparameters of the NN before overfitting sets in (see
    section~\ref{sec:finalAccuracy}), we obtain the \texttt{heterotic orbiencoder} of the input data. 
    The user can modify the specifications of the NN by using the explicit details
    provided in section~\ref{sec:HOE}.
    \label{fig:imagen_appendixA}}
\end{figure}

\end{landscape}

\section{Challenges to optimize accuracy and speed over overfitting}
\label{app:overfitting}

\begin{figure}[b!]
    \begin{center}
     \includegraphics[scale=0.2]{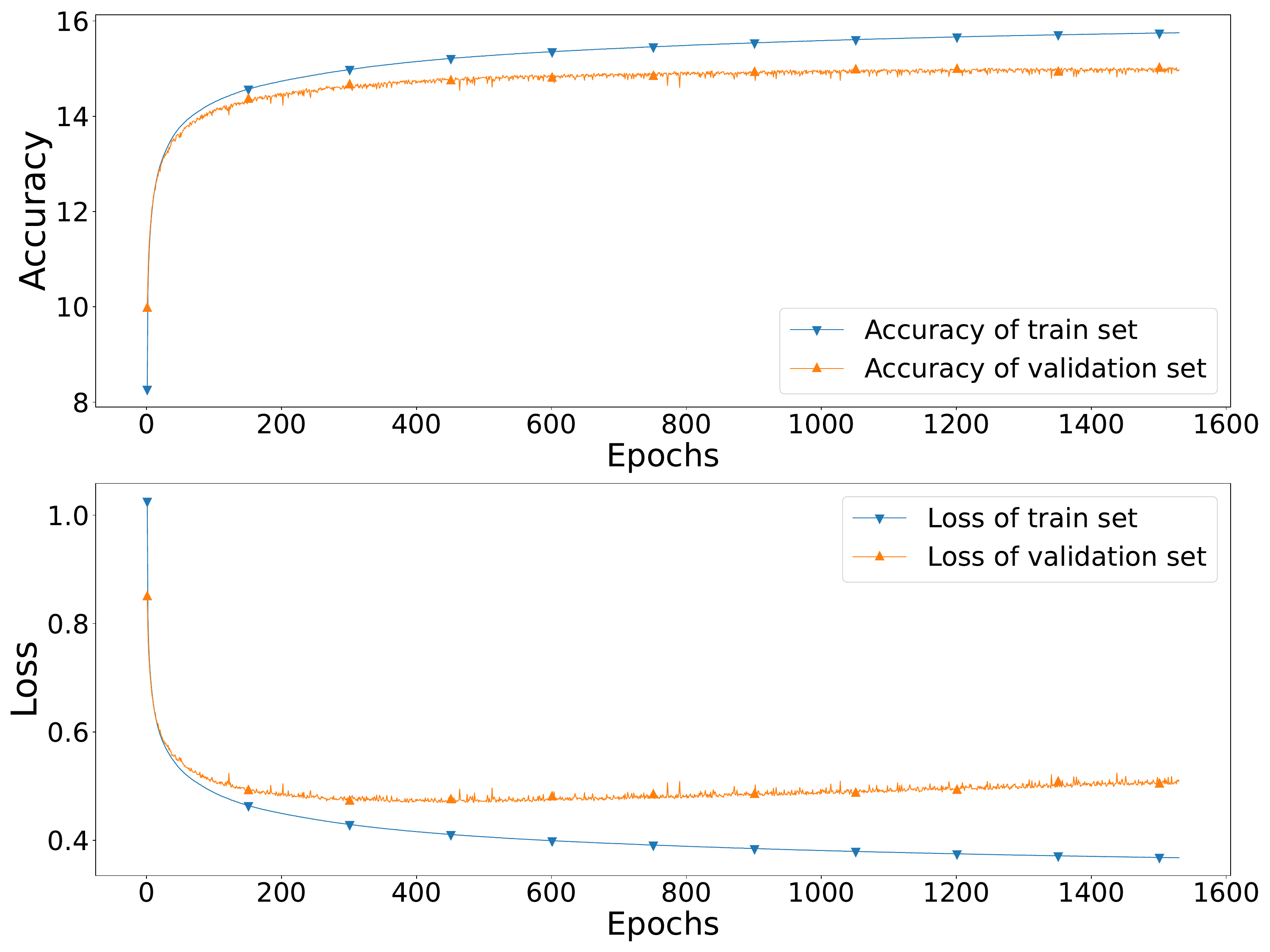}
    \end{center}
    \caption{Performance of the neural network trained with the parameters shown in 
    table~\ref{tab:finalconfig}, with an extended training time. We see 
    that the NN quickly arrives at great accuracy (about 15 features by 
    epoch 500). However, we also realize that the validation loss reaches
    a minimum around epoch 450 and then grows smoothly, indicating overfitting. 
    \label{fig:overfitting1500epochs}}
\end{figure}

Using our vanilla configuration with the default parameters and architecture 
replaced by those given in table~\ref{tab:finalconfig}, we have trained our 
autoencoder over 1,500 epochs, aiming at a high accuracy. We arrived at about 80\% 
accuracy (15.8 out of 20 features correctly predicted) in the training set 
and 75\% accuracy in the validation set, as shown in figure~\ref{fig:overfitting1500epochs}.
However, we realize that at epoch 448 the validation loss starts to
raise after reaching its minimum, while the training loss continues its
drop. This is a clear sign of overfitting, i.e.~the NN has found a way to
better characterize the models in the training set while becoming unable 
to characterize unseen models.

In order to overcome overfitting, we incorporated some popular proposals found in 
the literature. We explain here some of our efforts only for pedagogical 
reasons, for all trials returned negative results. We consider that the main 
observation from this discussion is that in our general orbifold autoencoder 
the best results are obtained by training a few epochs.

\begin{figure}[h!]
    \begin{center}
     \includegraphics[scale=0.2]{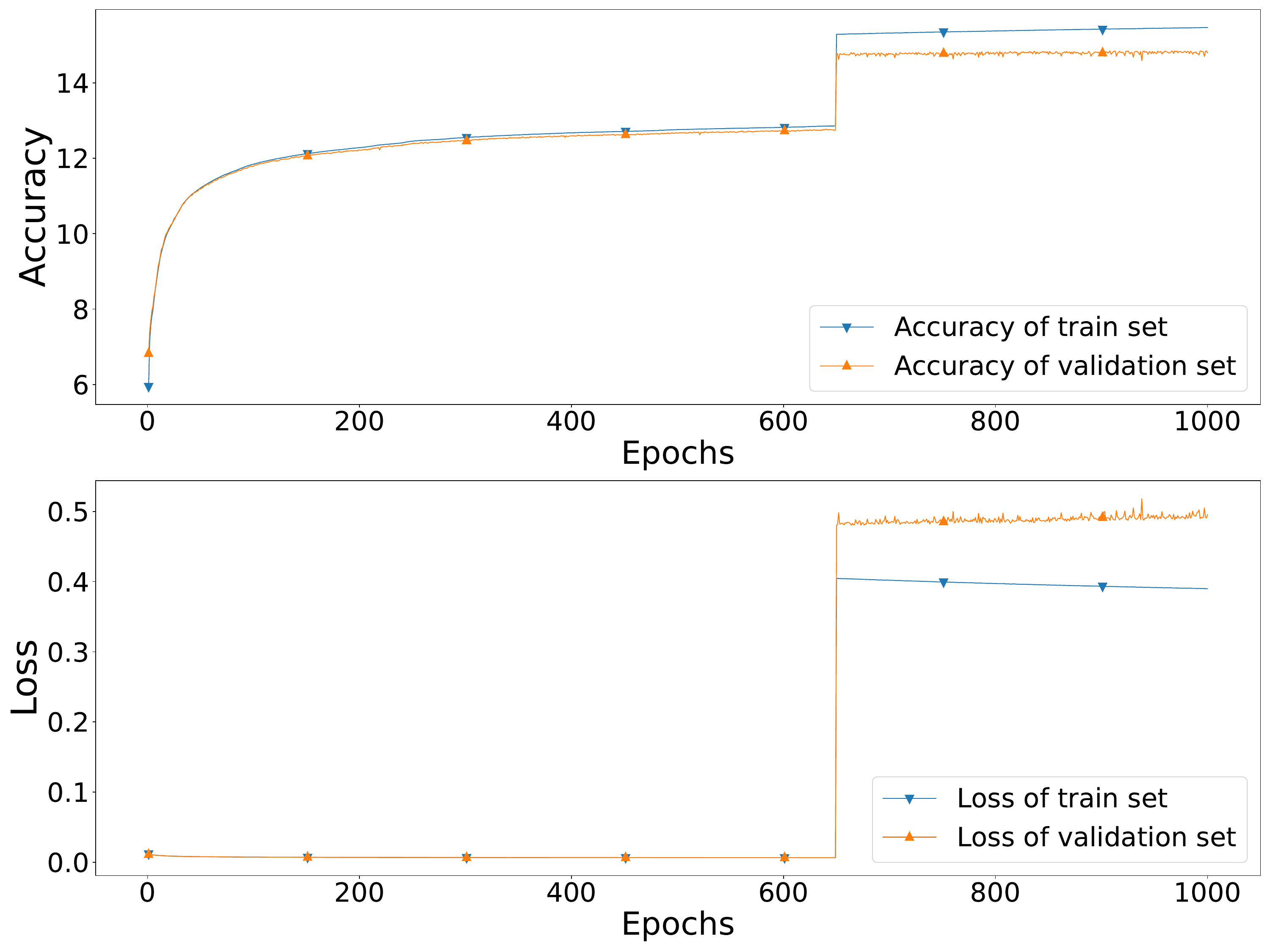}
    \end{center}
    \caption{This NN was trained with the parameters shown in 
    table~\ref{tab:finalconfig}, including additionally a scheduler to change 
    the learning rate, taking $0.1$ of the previous value, every 650 epochs. 
    We see that the NN accuracy jumps to a peak, but it is immediately 
    followed by overfitting. \label{figA1}}
\end{figure}

Our first approach was to introduce a scheduler, changing the learning rate 
every 650 epochs by reducing the learning rate to $0.1$ of the previous value. 
This approach is used to reduce the step size in the optimization algorithm, 
aiming at avoiding escaping from a given minimum, at a certain epoch. The 
results obtained after including the scheduler showed a quick improvement 
in accuracy, but were immediatly followed by overfitting and hence was not 
further applied, see figure~\ref{figA1}. 

\begin{figure}[h!]
    \begin{center}
     \includegraphics[scale=0.2]{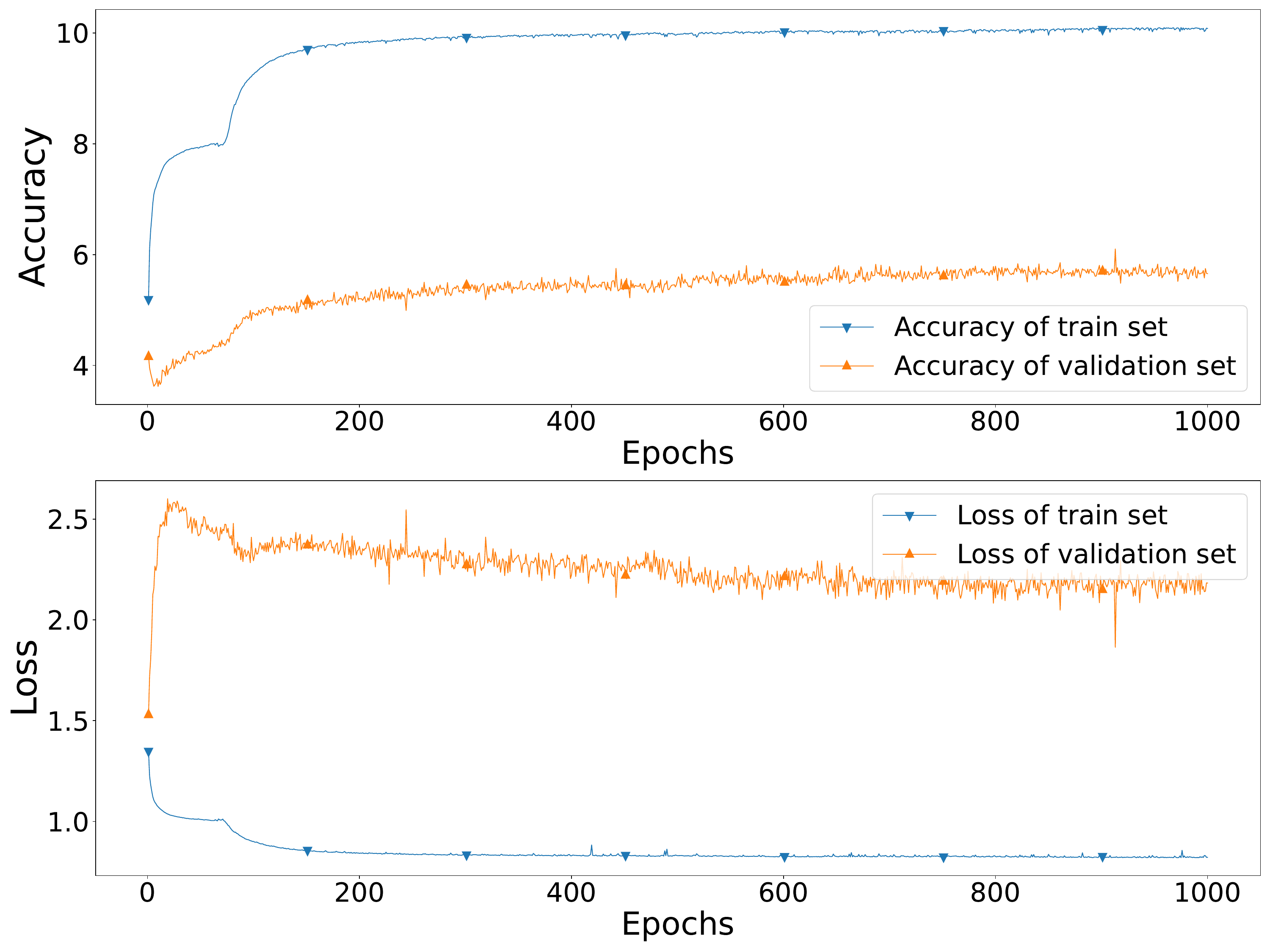}
    \end{center}
    \caption{This NN was trained with the parameters shown in table~\ref{tab:finalconfig}, 
    including now two 50\% dropout layers after the first and last hidden layers. 
    We observe that, although the overfitting is overcome, the accuracy is greatly 
    diminished, and the validation accuracy and loss behave erratically.
    \label{figA2}}
\end{figure}

Our second approach was to include dropouts in the first and last hidden 
layers, with 50\% of the neurons. As pointed in~\cite{https://doi.org/10.48550/arxiv.1207.0580},
these dropout phases help typically to reduce overfitting. Unfortunately, this 
also decreases dramatically the accuracy to approximately 10 out of 20 features 
at epoch 1,000. Extending the number of epochs did not increase the accuracy to 
acceptable values, see figure~\ref{figA2}.

\begin{figure}[t!]
    \begin{center}
     \includegraphics[scale=0.2]{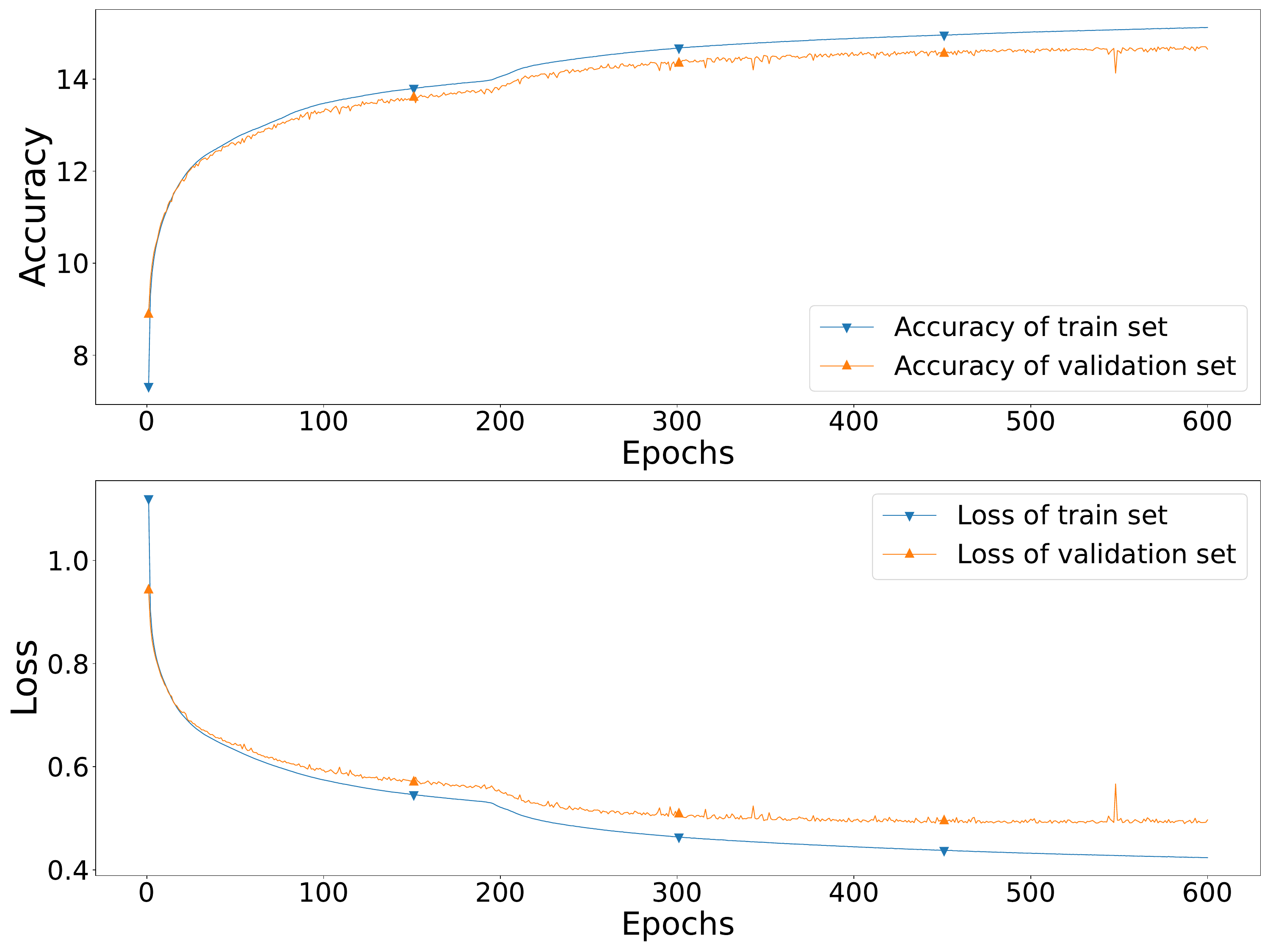}
    \end{center}
    \caption{We use the parameters of table \ref{tab:finalconfig} except for the 
    first and last hidden layers, which were reduced by 325 neurons each.  We see 
    that although the overfitting is overcome, the accuracy achieved at epoch 600 
    is similar to the one achieved prior to reducing the number of neurons. Hence, we
    find a similar behavior but with higher computation time.\label{figA3}}
\end{figure}

Our third attempt was to reduce the size of the first and last hidden layers, 
setting in 400 and 325 neurons less on those layers. We observe that although 
overfitting is overcome, the accuracy at epoch 1,000 is less than the accuracy 
of the original approach around epoch 400, and extending the training for a few 
extra epochs turns out in a huge computational effort with negligible increase 
in the accuracy, see figure~\ref{figA3}.

\section{\boldmath \SU5 GUTs from heterotic orbifolds\unboldmath}
\label{sec:SU5}

For illustration, we present the details of a sample \Z8 orbifold model leading to the spectrum
of the \SU5 GUT. Besides the parameters that define the \Z8--I (1,1) orbifold geometry
(see e.g.~\cite{GrootNibbelink:2013wnc,Olguin-Trejo:2019hxk}), in the notation of section 
\ref{sec:orbifolds} we need the parameters
\begin{subequations}
\begin{eqnarray}
  V_{1} = V_2 = V_3 = V_4 & = & 
        \left(\tfrac{1}{2}, \tfrac{1}{2}, \tfrac{1}{2}, \tfrac{1}{2}, \tfrac{1}{2}, \tfrac{1}{2},     1,     2\right)\left(-\tfrac{1}{2},    -2,     0,     0,     0,     1, -\tfrac{1}{2},     2\right),\\
  V_{5} = V_6 & = & \left(-\tfrac{3}{2},     0,     0, \tfrac{1}{2}, \tfrac{1}{2},     2, -\tfrac{3}{2},     1\right) \left(\tfrac{3}{2},     1, -\tfrac{1}{2}, \tfrac{1}{2}, \tfrac{3}{2},     0,    -1,    -2\right),\\
  V_7  & = & \left(\tfrac{1}{16}, \tfrac{1}{16}, \tfrac{1}{16}, \tfrac{1}{16}, \tfrac{1}{16}, \tfrac{1}{16}, \tfrac{1}{16}, \tfrac{1}{16}\right)  \left(-\tfrac{3}{8}, -\tfrac{1}{8},     0,     0,     0,     0, \tfrac{1}{8}, \tfrac{1}{8}\right),\\
  V_8  & = & 0\,.
\end{eqnarray}
\end{subequations}
By using the \texttt{orbifolder}~\cite{Nilles:2011aj}, we find that these parameters lead 
to the massless matter spectrum
\begin{equation}
\label{eq:Z8-GUTspectrum}
   3 \x\rep{10} \oplus 3 \x\crep{5} \oplus 3\x\rep1 \oplus 
   \left[5\x\rep5 \oplus 5\x\crep5 \oplus 217\x\rep1 \right]_\text{exotics}\,,
\end{equation}
which corresponds to the spectrum of an \SU5 GUT with three families,
including right-handed neutrinos, plus some vectorlike exotics (in parentheses)
that are decoupled at high energies~\cite{Lebedev:2007hv}.

This information is encoded in the 3D latent space of our autoencoder at the point
\begin{equation}
\label{eq:Z8SU5latentPoint}
  \{y_1, y_2, y_3\} ~=~ \{10.64, 9.27, 6.49\}\,.
\end{equation}

Analogously, we present a \Z{12}--I (1,1) orbifold model which, besides the 
parameters that fix the toroidal geometry of the compact space, requires the 
Wilson lines and shifts given by
\begin{subequations}
\begin{eqnarray}
  V_{1} ~=~ V_2 & = & V_3 ~=~ V_4 ~=~ 0\,,\\
  V_{5} ~=~ V_6 & = & \left(-\tfrac{1}{3},     0,    -1, -\tfrac{2}{3}, \tfrac{1}{3}, \tfrac{1}{3}, \tfrac{1}{3}, \tfrac{1}{3}\right) \left(-\tfrac{1}{3},     0,     0, \tfrac{1}{3}, \tfrac{4}{3}, -\tfrac{4}{3}, \tfrac{1}{3}, \tfrac{1}{3}\right),\\
  V_7  & = & \left(-\tfrac{5}{24}, -\tfrac{5}{24}, \tfrac{1}{24}, \tfrac{1}{24}, \tfrac{1}{24}, \tfrac{1}{24}, \tfrac{1}{24}, \tfrac{1}{24}\right)  \left(-\tfrac{1}{24}, -\tfrac{1}{24}, -\tfrac{1}{24}, -\tfrac{1}{24}, -\tfrac{1}{24}, \tfrac{1}{24}, \tfrac{1}{24}, \tfrac{1}{8}\right),\\
  V_8  & = & 0\,.
\end{eqnarray}
\end{subequations}
These parameters lead to the massless matter spectrum 
\begin{equation}
\label{eq:Z12-GUTspectrum}
   3 \x\rep{10} \oplus 3 \x\crep{5} \oplus 3\x\rep1 \oplus 
   \left[4\x\rep5 \oplus 4\x\crep5 \oplus 165\x\rep1 \right]_\text{exotics}\,,
\end{equation}
which is encoded in
\begin{equation}
\label{eq:Z12SU5latentPoint}
  \{y_1, y_2, y_3\} ~=~ \{10.39, 8.96, 9.07\}\,.
\end{equation}
Notice the great similarity of the points given in eq.~\eqref{eq:Z8SU5latentPoint}
and eq.~\eqref{eq:Z12SU5latentPoint}. This reveals the proximity of the most
promising models in the latent space.


{\small

\providecommand{\bysame}{\leavevmode\hbox to3em{\hrulefill}\thinspace}

}

\begin{thebibliography}{10}

\bibitem{He:2021kbj}
W.~He, T.~Li, R.~Sun, and L.~Wu, \emph{{The final model building for the
  supersymmetric Pati\textendash{}Salam models from intersecting D6-branes}},
  Eur. Phys. J. C \textbf{82} (2022), no.~8, 710, \texttt{arXiv:2112.09630}
  [hep-th].

\bibitem{He:2021gug}
W.~He, T.~Li, and R.~Sun, \emph{{The complete search for the supersymmetric
  Pati-Salam models from intersecting D6-branes}}, JHEP \textbf{08} (2022),
  044, \texttt{arXiv:2112.09632} [hep-th].

\bibitem{Loges:2022mao}
G.~J. Loges and G.~Shiu, \emph{{215 Billion Intersecting Brane Models}},
  (2022), \texttt{arXiv:2206.03506} [hep-th].

\bibitem{Senechal:1989me}
D.~Senechal, \emph{{Search for Four-dimensional String Models. 1.}}, Phys. Rev.
  D \textbf{39} (1989), 3717.

\bibitem{Dienes:1990ij}
K.~R. Dienes, \emph{{New string partition functions with vanishing cosmological
  constant}}, Phys. Rev. Lett. \textbf{65} (1990), 1979--1982.

\bibitem{Dienes:2006ut}
K.~R. Dienes, \emph{{Statistics on the heterotic landscape: Gauge groups and
  cosmological constants of four-dimensional heterotic strings}}, Phys. Rev. D
  \textbf{73} (2006), 106010, \texttt{hep-th/0602286}.

\bibitem{Dijkstra:2004cc}
T.~P.~T. Dijkstra, L.~R. Huiszoon, and A.~N. Schellekens, \emph{{Supersymmetric
  standard model spectra from RCFT orientifolds}}, Nucl. Phys. B \textbf{710}
  (2005), 3--57, \texttt{hep-th/0411129}.

\bibitem{Anastasopoulos:2006da}
P.~Anastasopoulos, T.~P.~T. Dijkstra, E.~Kiritsis, and A.~N. Schellekens,
  \emph{{Orientifolds, hypercharge embeddings and the Standard Model}}, Nucl.
  Phys. B \textbf{759} (2006), 83--146, \texttt{hep-th/0605226}.

\bibitem{Kreuzer:1998vb}
M.~Kreuzer and H.~Skarke, \emph{{Classification of reflexive polyhedra in
  three-dimensions}}, Adv. Theor. Math. Phys. \textbf{2} (1998), 853--871,
  \texttt{hep-th/9805190}.

\bibitem{Kreuzer:2000xy}
M.~Kreuzer and H.~Skarke, \emph{{Complete classification of reflexive polyhedra
  in four-dimensions}}, Adv. Theor. Math. Phys. \textbf{4} (2000), 1209--1230,
  \texttt{hep-th/0002240}.

\bibitem{Anderson:2011ns}
L.~B. Anderson, J.~Gray, A.~Lukas, and E.~Palti, \emph{{Two Hundred Heterotic
  Standard Models on Smooth Calabi-Yau Threefolds}}, Phys. Rev. D \textbf{84}
  (2011), 106005, \texttt{arXiv:1106.4804} [hep-th].

\bibitem{Gray:2012jy}
J.~Gray, Y.-H. He, V.~Jejjala, B.~Jurke, B.~D. Nelson, and J.~Simon,
  \emph{{Calabi-Yau Manifolds with Large Volume Vacua}}, Phys. Rev. D
  \textbf{86} (2012), 101901, \texttt{arXiv:1207.5801} [hep-th].

\bibitem{Scholler:2018apc}
F.~Sch\"oller and H.~Skarke, \emph{{All Weight Systems for
  Calabi\textendash{}Yau Fourfolds from Reflexive Polyhedra}}, Commun. Math.
  Phys. \textbf{372} (2019), no.~2, 657--678, \texttt{arXiv:1808.02422}
  [hep-th].

\bibitem{Lebedev:2006kn}
O.~Lebedev, H.~P. Nilles, S.~Raby, S.~Ramos-S{\'a}nchez, M.~Ratz, P.~K.~S.
  Vaudrevange, and A.~Wingerter, \emph{A mini-landscape of exact {MSSM} spectra
  in heterotic orbifolds}, Phys. Lett. \textbf{B645} (2007), 88,
  \texttt{hep-th/0611095}.

\bibitem{Lebedev:2007hv}
O.~Lebedev, H.~P. Nilles, S.~Raby, S.~Ramos-S{\'a}nchez, M.~Ratz, P.~K.~S.
  Vaudrevange, and A.~Wingerter, \emph{The heterotic road to the {MSSM} with
  {R} parity}, Phys. Rev. \textbf{D77} (2007), 046013, \texttt{arXiv:0708.2691
  [hep-th]}.

\bibitem{Lebedev:2008un}
O.~Lebedev, H.~P. Nilles, S.~Ramos-S\'{a}nchez, M.~Ratz, and P.~K.~S.
  Vaudrevange, \emph{{Heterotic mini-landscape (II): completing the search for
  MSSM vacua in a $Z_6$ orbifold}}, Phys. Lett. \textbf{B668} (2008), 331--335,
  \texttt{arXiv:0807.4384} [hep-th].

\bibitem{Nilles:2014owa}
H.~P. Nilles and P.~K.~S. Vaudrevange, \emph{{Geography of Fields in Extra
  Dimensions: String Theory Lessons for Particle Physics}}, Mod. Phys. Lett.
  \textbf{A30} (2015), no.~10, 1530008, \texttt{arXiv:1403.1597} [hep-th].

\bibitem{Olguin-Trejo:2018wpw}
Y.~Olgu{\'i}n-Trejo, R.~P{\'e}rez-Mart{\'i}nez, and S.~Ramos-S{\'a}nchez,
  \emph{{Charting the flavor landscape of MSSM-like Abelian heterotic
  orbifolds}}, Phys. Rev. \textbf{D98} (2018), no.~10, 106020,
  \texttt{arXiv:1808.06622} [hep-th].

\bibitem{Parr:2019bta}
E.~Parr and P.~K.~S. Vaudrevange, \emph{{Contrast data mining for the MSSM from
  strings}}, Nucl. Phys. \textbf{B952} (2020), 114922,
  \texttt{arXiv:1910.13473} [hep-th].

\bibitem{Blaszczyk:2014qoa}
M.~Blaszczyk, S.~Groot~Nibbelink, O.~Loukas, and S.~Ramos-S{\'a}nchez,
  \emph{{Non-supersymmetric heterotic model building}}, JHEP \textbf{10}
  (2014), 119, \texttt{arXiv:1407.6362} [hep-th].

\bibitem{Perez-Martinez:2021zjj}
R.~P{\'e}rez-Mart{\'i}nez, S.~Ramos-S{\'a}nchez, and P.~K.~S. Vaudrevange,
  \emph{{Landscape of promising nonsupersymmetric string models}}, Phys. Rev. D
  \textbf{104} (2021), no.~4, 046026, \texttt{arXiv:2105.03460} [hep-th].

\bibitem{Dixon:1985jw}
L.~J. Dixon, J.~A. Harvey, C.~Vafa, and E.~Witten, \emph{{Strings on
  Orbifolds}}, Nucl. Phys. B \textbf{261} (1985), 678--686.

\bibitem{Dixon:1986jc}
L.~J. Dixon, J.~A. Harvey, C.~Vafa, and E.~Witten, \emph{{Strings on Orbifolds.
  2.}}, Nucl. Phys. B \textbf{274} (1986), 285--314.

\bibitem{Ramos-Sanchez:2021woq}
S.~Ramos-S{\'a}nchez, M.~Ratz, Y.~Shirman, S.~Shukla, and M.~Waterbury,
  \emph{{Generation flow in field theory and strings}}, JHEP \textbf{10}
  (2021), 144, \texttt{arXiv:2109.01681} [hep-th].

\bibitem{Denef:2004ze}
F.~Denef and M.~R. Douglas, \emph{Distributions of flux vacua}, JHEP
  \textbf{05} (2004), 072, \texttt{hep-th/0404116}.

\bibitem{Giryavets:2004zr}
A.~Giryavets, S.~Kachru, and P.~K. Tripathy, \emph{On the taxonomy of flux
  vacua}, JHEP \textbf{08} (2004), 002, \texttt{hep-th/0404243}.

\bibitem{Gmeiner:2005vz}
F.~Gmeiner, R.~Blumenhagen, G.~Honecker, D.~L{\"u}st, and T.~Weigand, \emph{One
  in a billion: {MSSM}-like {D}-brane statistics}, JHEP \textbf{01} (2006),
  004, \texttt{hep-th/0510170}.

\bibitem{Douglas:2006xy}
M.~R. Douglas and W.~Taylor, \emph{The landscape of intersecting brane models},
   (2006), \texttt{hep-th/0606109}.

\bibitem{Abel:2014xta}
S.~Abel and J.~Rizos, \emph{{Genetic Algorithms and the Search for Viable
  String Vacua}}, JHEP \textbf{08} (2014), 010, \texttt{arXiv:1404.7359}
  [hep-th].

\bibitem{Cole:2019enn}
A.~Cole, A.~Schachner, and G.~Shiu, \emph{{Searching the Landscape of Flux
  Vacua with Genetic Algorithms}}, JHEP \textbf{11} (2019), 045,
  \texttt{arXiv:1907.10072} [hep-th].

\bibitem{Ruehle:2020jrk}
F.~Ruehle, \emph{Data science applications to string theory}, Phys. Rept.
  \textbf{839} (2020), 1--117.

\bibitem{Damian:2013dq}
C.~Dami{\'a}n, L.~R. D{\'i}az-Barr{\'o}n, O.~Loaiza-Brito, and M.~Sabido,
  \emph{{Slow-Roll Inflation in Non-geometric Flux Compactification}}, JHEP
  \textbf{06} (2013), 109, \texttt{arXiv:1302.0529} [hep-th].

\bibitem{Damian:2013dwa}
C.~Dami{\'a}n and O.~Loaiza-Brito, \emph{{More stable de Sitter vacua from
  S-dual nongeometric fluxes}}, Phys. Rev. D \textbf{88} (2013), no.~4, 046008,
  \texttt{arXiv:1304.0792} [hep-th].

\bibitem{Loges:2021hvn}
G.~J. Loges and G.~Shiu, \emph{{Breeding Realistic D-Brane Models}}, Fortsch.
  Phys. \textbf{70} (2022), no.~5, 2200038, \texttt{arXiv:2112.08391} [hep-th].

\bibitem{Carifio:2017bov}
J.~Carifio, J.~Halverson, D.~Krioukov, and B.~D. Nelson, \emph{{Machine
  Learning in the String Landscape}}, JHEP \textbf{09} (2017), 157,
  \texttt{arXiv:1707.00655} [hep-th].

\bibitem{Ashmore:2019wzb}
A.~Ashmore, Y.-H. He, and B.~A. Ovrut, \emph{{Machine Learning
  Calabi\textendash{}Yau Metrics}}, Fortsch. Phys. \textbf{68} (2020), no.~9,
  2000068, \texttt{arXiv:1910.08605} [hep-th].

\bibitem{He:2017aed}
Y.-H. He, \emph{Deep-learning the landscape},  (2017),
  \texttt{arXiv:1706.02714} [hep-th].

\bibitem{He:2019vsj}
Y.-H. He and S.-J. Lee, \emph{{Distinguishing elliptic fibrations with AI}},
  Phys. Lett. B \textbf{798} (2019), 134889, \texttt{arXiv:1904.08530}
  [hep-th].

\bibitem{Demirtas:2020dbm}
M.~Demirtas, L.~McAllister, and A.~Rios-Tascon, \emph{{Bounding the
  Kreuzer-Skarke Landscape}}, Fortsch. Phys. \textbf{68} (2020), 2000086,
  \texttt{arXiv:2008.01730} [hep-th].

\bibitem{Berglund:2021ztg}
P.~Berglund, B.~Campbell, and V.~Jejjala, \emph{Machine learning kreuzer-skarke
  calabi-yau threefolds},  (2021), \texttt{arXiv:2112.09117} [hep-th].

\bibitem{Jejjala:2022lxh}
V.~Jejjala, W.~Taylor, and A.~Turner, \emph{Identifying equivalent calabi--yau
  topologies: A discrete challenge from math and physics for machine learning},
   (2022), \texttt{arXiv:2202.07590} [hep-th].

\bibitem{Bull:2018uow}
K.~Bull, Y.-H. He, V.~Jejjala, and C.~Mishra, \emph{{Machine Learning CICY
  Threefolds}}, Phys. Lett. B \textbf{785} (2018), 65--72,
  \texttt{arXiv:1806.03121} [hep-th].

\bibitem{Erbin:2020tks}
H.~Erbin and R.~Finotello, \emph{{Machine learning for complete intersection
  Calabi-Yau manifolds: a methodological study}}, Phys. Rev. D \textbf{103}
  (2021), no.~12, 126014, \texttt{arXiv:2007.15706} [hep-th].

\bibitem{Halverson:2019tkf}
J.~Halverson, B.~Nelson, and F.~Ruehle, \emph{Branes with brains: Exploring
  string vacua with deep reinforcement learning}, JHEP \textbf{06} (2019), 003,
  \texttt{arXiv:1903.11616} [hep-th].

\bibitem{Constantin:2021for}
A.~Constantin, T.~R. Harvey, and A.~Lukas, \emph{Heterotic string model
  building with monad bundles and reinforcement learning},  (2021),
  \texttt{arXiv:2108.07316} [hep-th].

\bibitem{Abel:2021rrj}
S.~Abel, A.~Constantin, T.~R. Harvey, and A.~Lukas, \emph{Evolving heterotic
  gauge backgrounds: Genetic algorithms versus reinforcement learning},
  Fortsch. Phys. \textbf{70} (2022), no.~5, 2200034, \texttt{arXiv:2110.14029}
  [hep-th].

\bibitem{Mutter:2018sra}
A.~M\"utter, E.~Parr, and P.~K.~S. Vaudrevange, \emph{{Deep learning in the
  heterotic orbifold landscape}}, Nucl. Phys. B \textbf{940} (2019), 113--129,
  \texttt{arXiv:1811.05993} [hep-th].

\bibitem{Otsuka:2020nsk}
H.~Otsuka and K.~Takemoto, \emph{{Deep learning and k-means clustering in
  heterotic string vacua with line bundles}}, JHEP \textbf{05} (2020), 047,
  \texttt{arXiv:2003.11880} [hep-th].

\bibitem{Deen:2020dlf}
R.~Deen, Y.-H. He, S.-J. Lee, and A.~Lukas, \emph{{Machine learning string
  standard models}}, Phys. Rev. D \textbf{105} (2022), no.~4, 046001,
  \texttt{arXiv:2003.13339} [hep-th].

\bibitem{Halverson:2020opj}
J.~Halverson and C.~Long, \emph{{Statistical Predictions in String Theory and
  Deep Generative Models}}, Fortsch. Phys. \textbf{68} (2020), no.~5, 2000005,
  \texttt{arXiv:2001.00555} [hep-th].

\bibitem{CaboBizet:2020cse}
N.~Cabo~Bizet, C.~Dami{\'a}n, O.~Loaiza-Brito, D.~K. Mayorga~Pe{\~n}a, and
  J.~A. Monta{\~n}ez-Barrera, \emph{{Testing Swampland Conjectures with Machine
  Learning}}, Eur. Phys. J. C \textbf{80} (2020), no.~8, 766,
  \texttt{arXiv:2006.07290} [hep-th].

\bibitem{Damian:2022xfk}
C.~Dami{\'a}n and O.~Loaiza-Brito, \emph{{Metastable vacua from torsion and
  machine learning}},  (2022), \texttt{arXiv:2205.12373} [hep-th].

\bibitem{baldi2012autoencoders}
P.~Baldi, \emph{Autoencoders, unsupervised learning, and deep architectures},
  in \emph{Proceedings of ICML workshop on unsupervised and transfer learning},
  2012, pp.~37--49.

\bibitem{Fischer:2012qj}
M.~Fischer, M.~Ratz, J.~Torrado, and P.~K.~S. Vaudrevange,
  \emph{{Classification of symmetric toroidal orbifolds}}, JHEP \textbf{01}
  (2013), 084, \texttt{arXiv:1209.3906} [hep-th].

\bibitem{Ploger:2007iq}
F.~Pl{\"o}ger, S.~Ramos-S{\'a}nchez, M.~Ratz, and P.~K.~S. Vaudrevange,
  \emph{Mirage torsion}, JHEP \textbf{04} (2007), 063, \texttt{hep-th/0702176}.

\bibitem{Nilles:2011aj}
H.~P. Nilles, S.~Ramos-S{\'a}nchez, P.~K.~S. Vaudrevange, and A.~Wingerter,
  \emph{{The Orbifolder: A Tool to study the Low Energy Effective Theory of
  Heterotic Orbifolds}}, Comput. Phys. Commun. \textbf{183} (2012), 1363--1380,
  \texttt{arXiv:1110.5229} [hep-th].

\bibitem{Nibbelink:2013lua}
S.~Groot~Nibbelink and O.~Loukas, \emph{{MSSM-like models on {$\mathbbm{Z}_8$}
  toroidal orbifolds}}, JHEP \textbf{12} (2013), 044, \texttt{arXiv:1308.5145}
  [hep-th].

\bibitem{Olguin-Trejo:2019hxk}
Y.~Olgu{\'i}n-Trejo, O.~P{\'e}rez-Figueroa, R.~P{\'e}rez-Mart{\'i}nez, and
  S.~Ramos-S{\'a}nchez, \emph{{U(1)' coupling constant at low energies from
  heterotic orbifolds}}, Phys. Lett. B \textbf{795} (2019), 673--681,
  \texttt{arXiv:1901.10102} [hep-ph].

\bibitem{Kim:2006hw}
J.~E. Kim and B.~Kyae, \emph{{Flipped SU(5) from Z(12-I) orbifold with Wilson
  line}}, Nucl. Phys. B \textbf{770} (2007), 47--82, \texttt{hep-th/0608086}.

\bibitem{Kim:2007mt}
J.~E. Kim, J.-H. Kim, and B.~Kyae, \emph{{Superstring standard model from
  Z(12-I) orbifold compactification with and without exotics, and effective
  R-parity}}, JHEP \textbf{06} (2007), 034, \texttt{hep-ph/0702278}.

\bibitem{Huh:2009nh}
J.-H. Huh, J.~E. Kim, and B.~Kyae, \emph{{SU(5)(flip) x SU(5)-prime from
  Z(12-I)}}, Phys. Rev. D \textbf{80} (2009), 115012, \texttt{arXiv:0904.1108}
  [hep-ph].

\bibitem{Bailin:1999nk}
D.~Bailin and A.~Love, \emph{Orbifold compactifications of string theory},
  Phys. Rept. \textbf{315} (1999), 285--408.

\bibitem{Ramos-Sanchez:2008nwx}
S.~Ramos-S{\'a}nchez, \emph{{Towards Low Energy Physics from the Heterotic
  String}}, Fortsch. Phys. \textbf{10} (2009), 907--1036,
  \texttt{arXiv:0812.3560} [hep-th].

\bibitem{Vaudrevange:2008sm}
P.~K.~S. Vaudrevange, \emph{{Grand Unification in the Heterotic Brane World}},
  (2008), \texttt{arXiv:0812.3503} [hep-th].

\bibitem{Choi:2020dws}
K.-S. Choi and J.~E. Kim, \emph{{Quarks and Leptons From Orbifolded
  Superstring}}, Lect. Notes Phys. \textbf{954} (2020), pp.

\bibitem{GrootNibbelink:2017usl}
S.~Groot~Nibbelink and P.~K.~S. Vaudrevange, \emph{{T-duality orbifolds of
  heterotic Narain compactifications}}, JHEP \textbf{04} (2017), 030,
  \texttt{arXiv:1703.05323} [hep-th].

\bibitem{geometrystatistics}
E.~Escalante-Notario, I.~Portillo-Castillo, and S.~Ramos-S{\'a}nchez,
  \emph{Geometrystatistics}, 2022,
  \texttt{\url{https://github.com/enriqueescalante/makedataset/wiki/Sample-of-geometry-statistics}}.

\bibitem{Makedataset}
E.~Escalante-Notario, I.~Portillo-Castillo, and S.~Ramos-S{\'a}nchez,
  \emph{Makedataset}, 2022,
  \texttt{\url{https://github.com/enriqueescalante/makedataset}}.

\bibitem{glorot2010understanding}
X.~Glorot and Y.~Bengio, \emph{{Understanding the difficulty of training deep
  feedforward neural networks}}, in \emph{Proceedings of the thirteenth
  international conference on artificial intelligence and statistics}, JMLR
  Workshop and Conference Proceedings, 2010, pp.~249--256.

\bibitem{kumar2017weight}
S.~K. Kumar, \emph{{On weight initialization in deep neural networks}}, arXiv
  preprint arXiv:1704.08863 (2017).

\bibitem{koturwar2017weight}
S.~Koturwar and S.~Merchant, \emph{{Weight initialization of deep neural
  networks (DNNs) using data statistics}}, arXiv preprint arXiv:1710.10570
  (2017).

\bibitem{datta2020survey}
L.~Datta, \emph{{A survey on activation functions and their relation with
  xavier and he normal initialization}}, arXiv preprint arXiv:2004.06632
  (2020).

\bibitem{narkhede2022review}
M.~V. Narkhede, P.~P. Bartakke, and M.~S. Sutaone, \emph{{A review on weight
  initialization strategies for neural networks}}, Artificial intelligence
  review \textbf{55} (2022), no.~1, 291--322.

\bibitem{clevert2015fast}
D.-A. Clevert, T.~Unterthiner, and S.~Hochreiter, \emph{{Fast and accurate deep
  network learning by exponential linear units (elus)}}, arXiv preprint
  arXiv:1511.07289 (2015).

\bibitem{klambauer2017self}
G.~Klambauer, T.~Unterthiner, A.~Mayr, and S.~Hochreiter,
  \emph{Self-normalizing neural networks}, Advances in neural information
  processing systems \textbf{30} (2017).

\bibitem{pedamonti2018comparison}
D.~Pedamonti, \emph{{Comparison of non-linear activation functions for deep
  neural networks on MNIST classification task}}, arXiv preprint
  arXiv:1804.02763 (2018).

\bibitem{hayou2019impact}
S.~Hayou, A.~Doucet, and J.~Rousseau, \emph{On the impact of the activation
  function on deep neural networks training}, in \emph{International conference
  on machine learning}, PMLR, 2019, pp.~2672--2680.

\bibitem{dubey2021comprehensive}
S.~R. Dubey, S.~K. Singh, and B.~B. Chaudhuri, \emph{{A comprehensive survey
  and performance analysis of activation functions in deep learning}}, arXiv
  preprint arXiv:2109.14545 (2021).

\bibitem{janocha2017loss}
K.~Janocha and W.~M. Czarnecki, \emph{On loss functions for deep neural
  networks in classification}, arXiv preprint arXiv:1702.05659 (2017).

\bibitem{nie2018investigation}
F.~Nie, Z.~Hu, and X.~Li, \emph{An investigation for loss functions widely used
  in machine learning}, Communications in Information and Systems \textbf{18}
  (2018), no.~1, 37--52.

\bibitem{kingma2014adam}
D.~P. Kingma and J.~Ba, \emph{Adam: A method for stochastic optimization},
  arXiv preprint arXiv:1412.6980 (2014).

\bibitem{wilson2017marginal}
A.~C. Wilson, R.~Roelofs, M.~Stern, N.~Srebro, and B.~Recht, \emph{The marginal
  value of adaptive gradient methods in machine learning}, Advances in neural
  information processing systems \textbf{30} (2017).

\bibitem{choi2019empirical}
D.~Choi, C.~J. Shallue, Z.~Nado, J.~Lee, C.~J. Maddison, and G.~E. Dahl,
  \emph{On empirical comparisons of optimizers for deep learning}, arXiv
  preprint arXiv:1910.05446 (2019).

\bibitem{schneider2019deepobs}
F.~Schneider, L.~Balles, and P.~Hennig, \emph{Deepobs: A deep learning
  optimizer benchmark suite}, arXiv preprint arXiv:1903.05499 (2019).

\bibitem{GrootNibbelink:2013wnc}
S.~Groot~Nibbelink and O.~Loukas, \emph{{MSSM-like models on Z(8) toroidal
  orbifolds}}, JHEP \textbf{12} (2013), 044, \texttt{arXiv:1308.5145} [hep-th].

\bibitem{Kim:2011yz}
J.~E. Kim, \emph{{The CKM matrix with maximal CP violation from Z(12)
  symmetry}}, Phys. Lett. B \textbf{704} (2011), 360--366,
  \texttt{arXiv:1109.0995} [hep-ph].

\bibitem{Kim:2015mpa}
J.~E. Kim, \emph{{Towards unity of families: anti-SU(7) from $Z_{12-I}$
  orbifold compactification}}, JHEP \textbf{06} (2015), 114,
  \texttt{arXiv:1503.03104} [hep-ph].

\bibitem{Kim:2017ylg}
J.~E. Kim, B.~Kyae, and S.~Nam, \emph{{The anomalous $U(1)_{\mathrm{anom}}$
  symmetry and flavors from an $SU(5)\times SU(5)'$ GUT in $Z_{12-I}$ orbifold
  compactification}}, Eur. Phys. J. C \textbf{77} (2017), no.~12, 847,
  \texttt{arXiv:1703.05345} [hep-ph].

\bibitem{Jeong:2019zlr}
J.~Jeong, J.~E. Kim, and S.~Nam, \emph{{Leptonic CP violation in flipped SU(5)
  GUT from $Z_{12-I}$ orbifold compactification}}, Phys. Lett. B \textbf{791}
  (2019), 215--224, \texttt{arXiv:1901.02295} [hep-ph].

\bibitem{NEURIPS2019_9015}
A.~Paszke, S.~Gross, F.~Massa, A.~Lerer, J.~Bradbury, G.~Chanan, T.~Killeen,
  Z.~Lin, N.~Gimelshein, L.~Antiga, A.~Desmaison, A.~Kopf, E.~Yang, Z.~DeVito,
  M.~Raison, A.~Tejani, S.~Chilamkurthy, B.~Steiner, L.~Fang, J.~Bai, and
  S.~Chintala, \emph{Pytorch: An imperative style, high-performance deep
  learning library}, Advances in Neural Information Processing Systems 32,
  Curran Associates, Inc., 2019, pp.~8024--8035,
  \texttt{http://papers.neurips.cc/paper/9015-pytorch-an-imperative-style-high-performance-
  deep-learning-library.pdf}.

\bibitem{mckinney-proc-scipy-2010}
W.~McKinney, \emph{Data structures for statistical computing in python}, in
  \emph{Proceedings of the 9th Python in Science Conference} (S.~van~der Walt
  and J.~Millman, eds.), 2010, pp.~51 -- 56.

\bibitem{scikit-learn}
F.~Pedregosa, G.~Varoquaux, A.~Gramfort, V.~Michel, B.~Thirion, O.~Grisel,
  M.~Blondel, P.~Prettenhofer, R.~Weiss, V.~Dubourg, J.~Vanderplas, A.~Passos,
  D.~Cournapeau, M.~Brucher, M.~Perrot, and E.~Duchesnay, \emph{Scikit-learn:
  Machine learning in {P}ython}, Journal of Machine Learning Research
  \textbf{12} (2011), 2825--2830.

\bibitem{pmlr-v15-glorot11a}
X.~Glorot, A.~Bordes, and Y.~Bengio, \emph{Deep sparse rectifier neural
  networks}, in \emph{Proceedings of the Fourteenth International Conference on
  Artificial Intelligence and Statistics} (Fort Lauderdale, FL, USA)
  (G.~Gordon, D.~Dunson, and M.~Dudík, eds.), Proceedings of Machine Learning
  Research, vol.~15, PMLR, 11--13 Apr 2011, pp.~315--323,
  \texttt{https://proceedings.mlr.press/v15/glorot11a.html}.

\bibitem{DBLP:journals/corr/abs-1901-00434}
P.~Baldi and R.~Vershynin, \emph{The capacity of feedforward neural networks},
  \textbf{abs/1901.00434} (2019), \texttt{1901.00434},
  \texttt{http://arxiv.org/abs/1901.00434}.

\bibitem{maas2013rectifier}
A.~L. Maas, A.~Y. Hannun, A.~Y. Ng, et~al., \emph{Rectifier nonlinearities
  improve neural network acoustic models}, in \emph{Proc. icml}, vol.~30,
  Citeseer, 2013, p.~3.

\bibitem{AutoencoderST}
E.~Escalante-Notario, I.~Portillo-Castillo, and S.~Ramos-S{\'a}nchez,
  \emph{{Heterotic Orbiencoder}}, 2022,
  \texttt{\url{https://github.com/enriqueescalante/heterotic_orbiencoder}}.

\bibitem{Georgi:1974sy}
H.~Georgi and S.~L. Glashow, \emph{Unity of all elementary particle forces},
  Phys. Rev. Lett. \textbf{32} (1974), 438--441.

\bibitem{robustnesstests}
E.~Escalante-Notario, I.~Portillo-Castillo, and S.~Ramos-S{\'a}nchez,
  \emph{{RobustnessTests}}, 2022,
  \texttt{\url{https://github.com/enriqueescalante/heterotic_orbiencoder/wiki/Consistency-tests}}.

\bibitem{https://doi.org/10.48550/arxiv.1207.0580}
G.~E. Hinton, N.~Srivastava, A.~Krizhevsky, I.~Sutskever, and R.~R.
  Salakhutdinov, \emph{Improving neural networks by preventing co-adaptation of
  feature detectors},  (2012), \texttt{arXiv:1207.0580} [cs.NE].

\end{thebibliography}
\end{document}